\documentclass[twocolumn,superscriptaddress,aps,prx,amsmath,nofootinbib,amssymb,floatfix]{revtex4-2}
\usepackage{graphicx}
\usepackage{dcolumn}
\usepackage{bm}
\usepackage{dsfont}
\usepackage[breaklinks=true,colorlinks,citecolor=blue,linkcolor=blue,urlcolor=blue]{hyperref}

\usepackage{physics}
\usepackage{mleftright}
\usepackage{bbm}

\newcommand{\be}{\begin{equation}}
\newcommand{\ee}{\end{equation}}
\newcommand{\bea}{\begin{eqnarray}}
\newcommand{\eea}{\end{eqnarray}}

\renewcommand{\eqref}[1]{\mbox{Eq.~(\ref{#1})}}

\newcommand{\figpanel}[2]{Fig.~\hyperref[#1]{\ref*{#1}(#2)}}

\usepackage{changes}
\definechangesauthor[name={Ale}, color=red]{Ale}
\definechangesauthor[name={David}, color=orange]{David}

\usepackage{tikz,xcolor}
\definecolor{lime}{HTML}{A6CE39}
\DeclareRobustCommand{\orcidicon}{%
	\begin{tikzpicture}
	\draw[lime, fill=lime] (0,0) 
	circle [radius=0.16] 
	node[white] {{\fontfamily{qag}\selectfont \tiny ID}};
	\draw[white, fill=white] (-0.0625,0.095) 
	circle [radius=0.007];
	\end{tikzpicture}
	\hspace{-2mm}
}
\foreach \x in {A, ..., Z}{%
	\expandafter\xdef\csname orcid\x\endcsname{\noexpand\href{https://orcid.org/\csname orcidauthor\x\endcsname}{\noexpand\orcidicon}}
}

\begin{document}
\title{High-order interactions in quantum optomechanics: fluctuations, dynamics and thermodynamics}

\date{\today}

\author{Alessandro Ferreri\orcidA{}}
\affiliation{Institute for Quantum Computing Analytics (PGI-12), Forschungszentrum J\"ulich, 52425 J\"ulich, Germany}
\author{Vincenzo Macrì\orcidE{}}
\affiliation{Dipartimento di Fisica, Università di Pavia, Via Bassi 6, 27100 Pavia, Italy}
\author{Yoshihiko Hasegawa\orcidD{}}
\affiliation{Department of Information and Communication Engineering, Graduate School of Information Science and Technology, The University of Tokyo}
\author{David Edward Bruschi\orcidB{}}
\affiliation{Institute for Quantum Computing Analytics (PGI-12), 
Forschungszentrum J\"ulich, 52425 J\"ulich, Germany}
\affiliation{Theoretical Physics, Universit\"at des Saarlandes, 66123 Saarbr\"ucken, Germany}

\begin{abstract}
Quantum optomechanics describes the interaction between a confined field and a fluctuating wall due to radiation pressure. The dynamics of this system is typically understood using perturbation theory up to second order in the small coupling. Improving beyond this regime can shed light onto new phenomena.
In this work we study high-order resonant wall-field interactions characterized by  two- and three-phonon scattering processes. We obtain the Hamiltonian, compute the perturbed energy spectrum and explicitly calculate corrections to the ground state. Finally, we study the dynamics of the system when second- and third-order resonance conditions are activated, showing that the presence of high-order terms in the Hamiltonian drastically affects the populations of all particles, as well as the entropy production rate.

\end{abstract}

\maketitle

\section{Introduction}
Quantum optomechanics studies the interactions between the electromagnetic field and macroscopic objects, normally characterized by mechanical bosonic degrees of freedom \cite{meystreShortWalkQuantum2013,qvarfortTimeevolutionNonlinearOptomechanical2020}. In cavity optomechanics, mirrors with a vibrational degree of freedom confine the electromagnetic field and are modelled as quantum mechanical modes  \cite{kippenbergCavityOptomechanicsBackAction2008,aspelmeyerCavityOptomechanics2014}. This occurs, for example, when a cavity wall is allowed to fluctuate around an average given position \cite{lawInteractionMovingMirror1995,PhysRevA.106.033502,girvin2florian,bruschiTimeEvolutionCoupled2019}. At low temperatures, the vibrations of the wall essentially arise due to the quantum nature of the zero-point fluctuations \cite{marquardt2008quantum}, suggesting that the mirror effectively behaves as a quantum harmonic oscillator.

The simplest theoretical model to describe the coupling between the electromagnetic field and a quantized moving mirror in is absed on two bosonic modes interacting via radiation pressure effects \cite{bosePreparationNonclassicalStates1997,Nunnenkamp2011,meystreShortWalkQuantum2013,aspelmeyerCavityOptomechanics2014,heQuantumOptomechanicsLinearization2012}. Radiation pressure modifies the position of the mirror by means of photons inside the cavity impinging on, and being reflected by, the wall \cite{PhysRevLett.51.1550,cohadonCoolingMirrorRadiation1999}. Interestingly, this compelling model boasts several mathematical advantages. Among many, we mention here that the standard optomechanical Hamiltonian can be diagonalized analytically \cite{bowen2015quantum,clerk2014basic}, and that its unitary time evolution may be in general solved non-perturbatively \cite{bruschiMechanoopticsOptomechanicalQuantum2018,bruschiTimeEvolutionCoupled2019,qvarfortTimeevolutionNonlinearOptomechanical2020,schneiterOptimalEstimationQuantum2020}. Furthermore, the dynamics can be linearized assuming that the cavity mode is pumped coherently by the presence of a strong laser \cite{meystreShortWalkQuantum2013,aspelmeyerCavityOptomechanics2014,wollmanQuantumSqueezingMotion2015,PhysRevLett.112.150602}.

The validity of the above-mentioned model relies on two fundamental prerequisites: (i) the oscillation amplitude of any mechanical mode is much smaller than the cavity length, and (ii) the mechanical frequencies are much smaller than all optical frequencies \cite{meystreShortWalkQuantum2013}. The first assumption guarantees the absence of nonlinearities caused by broad oscillations of the wall. The second assumption legitimises the rotating wave approximation (RWA), which allows us to ignore both secularities and phonon-photon scattering terms in the Hamiltonian \cite{lawInteractionMovingMirror1995}. 

Although these two assumptions are reasonably justified by the state-of-the-art technology of experimental setups \cite{liuProgressOptomechanicalMicro2021,barzanjehOptomechanicsQuantumTechnologies2022}, recent literature has addressed optomechanical systems in a more extended fashion, namely by including within the Hamiltonian both phonon-photon scattering terms \cite{lawInteractionMovingMirror1995,armataVacuumEnergyDensities2015,armataNonequilibriumDressingCavity2017, PhysRevA.99.053815,10.21468/SciPostPhys.18.2.067} and quadratic terms in the mechanical oscillation amplitudes \cite{meystreShortWalkQuantum2013,liaoSinglephotonQuadraticOptomechanics2015,bruschiMechanoopticsOptomechanicalQuantum2018,PhysRevA.111.043524}. Interestingly, this extended formulation of the optomechanical models offers new perspectives and leads to the prediction of new intriguing effects. For example, it has already been shown that the inclusion of scattering terms in the Hamiltonian leads to the prediction phenomena such as vacuum Casimir-Rabi oscillations \cite{dodonov2005dynamical, macriNonperturbativeDynamicalCasimir2018}, as well as the conversion of incoherent phonons into correlated photon pairs \cite{settineriConversionMechanicalNoise2019}. Notably, the latter has paved the way for the conception of quantum heat engines based on a Casimir-Otto cycle \cite{PhysRevResearch.5.043274}. It is relevant to notice that the significant advancements in the realization of both high-frequency mechanical oscillators \cite{renningerBulkCrystallineOptomechanics2018, kharelHighfrequencyCavityOptomechanics2019,kharelMultimodeStrongCoupling2022,Diamandi:2024hpz} and quantum platforms simulating the dynamical Casimir effect \cite{wilsonObservationDynamicalCasimir2011,nationColloquiumStimulatingUncertainty2012,lahteenmakiDynamicalCasimirEffect2013,schneiderObservationBroadbandEntanglement2020} make the observations of these effects promising. We stress that heat transfers mediated by second-order effective wall-wall interactions have already been observed experimentally \cite{fongPhononHeatTransfer2019}.

In this work, we address the impact of second- and third-order Hamiltonian terms on both the spectrum, the dynamics and the thermodynamics of the system. We define first-order interactions as the Hamiltonian terms that are proportional to the oscillation amplitude $\delta L$ of the mirror, whereas second- and third-order Hamiltonian terms encode the nonlinear interactions between  wall field whose strength is proportional to $\delta L^2$ and $\delta L^3$ respectively. 

We first employ the formalism developed in the literature \cite{PhysRevA.106.033502} to compute the Hamiltonian of a scalar field confined in a 3-dimensional cavity with possesses a movable wall using perturbation theory up to the third order in $\delta L$. 
We then focus on the first two optical modes and use the Hamiltonian to calculate the eigenenergies of the system. 

Interestingly, we observe that the inclusion of high-order terms in the Hamiltonian strongly alter the energy levels in proximity of the energy splits, especially where the split is caused by second and third-order interactions. This suggests that high-order terms modify the effective coupling between field and wall.

Finally, we are interested in studying the dynamics and the thermodynamical properties of the system when mechanical and optical frequencies fulfill specific resonant conditions, while the only source of energy is a hot bath coupled to the wall. To do this, we employ the formalism of the generalized master equation, see \cite{settineriDissipationThermalNoise2018}, to numerically examine the bosonic populations, the heat-flow rates between each subsystem and the bath, as well as the entropy production rate. 
We observe that the role of high-order terms strongly depends on both the coupling strength between the two subsystems and the specific resonance condition we impose. Finally, we evaluate both the phononic population and the entropy production rate and find that they are particularly sensitive to the effective coupling between field and wall.

The paper is structured as follows: in Section \ref{model} we introduce the Hamiltonian of the 3-D box, from which we also acquire the Hamiltonian of the 1-D cavity. By means of the perturbation theory, we calculate the correction to the ground state caused by the interaction. Finally, in Section \ref{NA} we present our numerical analysis, in which we study the time evolution of populations, heat rates and entropy production rates when specific resonance conditions are activated.

\section{Highly nonlinear optomechanical model}\label{model}
\subsection{Hamiltonian of the 3-D and the 1-D model}
The system consists of a 3-dimensional cavity with volume $V=L_\text{x} L_\text{y} L_\text{z}$ that confines a massless scalar field. One of the walls of the cavity is allowed to oscillate around a fixed position $L_\text{x}$ with small oscillations. From now on, we will conveniently refer to the fluctuation amplitude of the wall via the adimensional amplitude $\epsilon:=\delta L_\text{x}/L_\text{x}\ll1$. 

To obtain the Hamiltonian of such system we apply the protocol described in \cite{PhysRevA.106.033502}. Such protocol is reported in Appendix \ref{H3app}.
This yields the Hamiltonian
\begin{align}\label{quantum:hamiltonian:explicit}
\hat{H}=&\hat H_0+\epsilon\hat H_{1}+\epsilon^2\hat H_{2}+\epsilon^3\hat H_{3},
\end{align}
where $\hat{H}_0:=\sum_{\bold n}\omega_{\bold n}\,(\hat{a}_{\bold n}^\dag\hat{a}_{\bold n}+\frac{1}{2})+\Omega(\hat{b}^\dag\hat{b}+\frac{1}{2})$ is the free Hamiltonian and the interaction terms read
\begin{align}\label{quantum:hamiltonian:terms2}
\hat{H}_{1}:=&2\sum_{\bold n}\left[\frac{k_{n_\perp}^2}{\omega_{\bold n}}\left(\hat a_{\bold n}^\dag\hat a_{\bold n}+\frac{1}{2}\right)\right.\nonumber\\
&\left.-2\sum_{m_x}(-1)^{n_\textrm{x}+m_\textrm{x}}\frac{k_{{n_\textrm{x}}}k_{{m_\textrm{x}}}}{\sqrt{\omega_{\bold n}\,\omega_{\bold m}}} \hat X_{\bold n}\hat X_{\bold m}\right]\hat X_{\textrm{b}},\nonumber\\
\hat{H}_{2}:=&-2\sum_{\bold n}\left[\frac{k_{{n_\textrm{x}}}^2k_{n_\perp}^2}{\omega_{\bold n}^3}\hat P_{\bold n}^2\right.\nonumber\\
&\left.+2\sum_{m_x}(-1)^{m_\textrm{x}+n_\textrm{x}}\frac{k_{{m_\textrm{x}}}k_{{n_\textrm{x}}}}{(\omega_{\bold n}\omega_{\bold m})^\frac{5}{2}}\zeta_{\bold n \bold m}\hat X_{\bold n}\hat X_{\bold m}\right]\hat X_{\textrm{b}}^2,\nonumber\\
\hat{H}_{3}:=&\sum_{\bold n}\sum_{m_x}(-1)^{m_\textrm{x}+n_\textrm{x}}\frac{k_{{m_\textrm{x}}}k_{{n_\textrm{x}}}}{6(\omega_{\bold n}\omega_{\bold m})^\frac{5}{2}}\left[\theta_{\bold n \bold m} (\hat a_{\bold n}^\dag \hat a_{\bold m}+ \hat a_{\bold n}\hat a_{\bold m}^\dag)\right.\nonumber\\
&\left.+\chi_{\bold n \bold m}(\hat a_{\bold n} \hat a_{\bold m}+\hat a_{\bold m}^\dag \hat a_{\bold n}^\dag)\right]\hat X_{\textrm{b}}^3.
\end{align}
In the formulas above $\Omega$ indicates the oscillation frequency of the wall, whereas the dispersion law of the scalar field reads $\omega_{\bold n}:=\sqrt{k_{{n_\textrm{x}}}^2+k_{n_\perp}^2}$,
where we define the wave vectors $ k_{{n_\textrm{x}}}=\pi n_\textrm{x}/L_\textrm{x}$, and $k_{n_\perp}^2=k_{{n_\textrm{y}}}^2+k_{{n_\textrm{z}}}^2=\left(\pi n_\textrm{y}/L_\textrm{y}\right)^2+\left(\pi n_\textrm{z}/L_\textrm{z}\right)^2$.
Quantum operators fulfill the canonical commutation relations $[\hat a_{\bold n},\hat a_{{\bold m}}^\dag]=\delta_{{\bold n}{\bold m}}$ for the field and
$[\hat{b},\hat{b}^\dag]=1$ for the harmonic oscillator.
We introduced the quadrature position and momentum operators $\hat X_{\textrm{b}}=\frac{1}{2}(\hat b^\dagger+\hat b)$, $\hat P_{\textrm{b}}=\frac{i}{2}(\hat b^\dagger-\hat b)$, $\hat X_{n}=\frac{1}{2}(\hat a_n^\dagger+\hat a_n)$, and $\hat P_{n}=\frac{i}{2}(\hat a_n^\dagger-\hat a_n)$. In our notation $\bold n\equiv(n_x,n_\text{y},n_\text{z})$ and $\bold m\equiv(m_x,n_\text{y},n_\text{z})$ (note that the two sets of indices only differ in the first integer). Each sum over $\bold n$ is therefore defined as $\sum_{\bold n}=\sum_{n_x}\sum_{n_y}\sum_{n_z}$. The explicit expression for $\zeta_{\bold n \bold m}$, $\theta_{\bold n \bold m}$ and $\chi_{\bold n \bold m}$ are reported in Appendix \ref{H3app}.

We recall that the first order of such Hamiltonian has already been predicted in \cite{PhysRevResearch.5.043274}, however, we want to spend a few words on the role of the first term in $\hat H_1$. First, we notice that such term does not contribute to the phonon-photon conversion, as the presence of wave vector components $k_{n_\perp}$ would not fulfill the momentum conservation in the direction y and z (phonons are created only along x), but it only affects the so-called \emph{radiation pressure term}. This reads
\begin{align}
\hat H_{\textrm{RP}}=2\sum_{\bold{n}}\frac{k_{n_\perp}^2-k_{n_x}^2}{\omega_{\bold n}}\left(\hat a_{\bold n}^\dag\hat a_{\bold n}+\frac{1}{2}\right)\hat X_\textrm{b}.
\end{align}
To provide a physical origin of this term, we note that inside the cavity each photon must have components in each direction due to the Heisenberg principle: the bounded space of the box increases the probability to find the photon in the box volume, thereby also increasing the uncertainty of the photon momentum. On the contrary, outside the box the field modes are not spatially bound and virtual photons pressing on the wall can safely possess only the component along x, as the infinite space outside can ensure high moment precision. Therefore, the presence of terms proportional to $k_{n_\perp}$ stems from the difference between the force pushing the wall from the inside the box, caused by photons whose momentum is subject to high uncertainty, as well as the force caused by photons pressing from outside.

The dynamics induced by the interaction Hamiltonian $\hat H_1$ have been extensively studied \cite{armataNonequilibriumDressingCavity2017, macriNonperturbativeDynamicalCasimir2018, PhysRevResearch.5.043274, PhysRevResearch.6.023320}. Therefore, here we focus on the dynamics induced by including the terms $\hat H_2$ and $\hat H_3$. We note that these terms encode four types of interaction:
\begin{itemize}
    \item \textit{Resonant interactions.} In the interaction Hamiltonians $\hat H_2$ and $\hat H_3$ one finds resonant interactions, which consist of four-wave-mixing-like phenomena described by terms of the form $\hat a_{\bold n}^\dag\hat a_{\bold m}^\dag\hat b^2+\textrm{h.c.}$ or $\hat a_{\bold n}^\dag\hat a_{\bold m}\hat b^2+\textrm{h.c.}$ and $\hat a_{\bold n}^\dag\hat a_{\bold m}^\dag\hat b^3+\textrm{h.c.}$ or $\hat a_{\bold n}^\dag\hat a_{\bold m}\hat b^3+\textrm{h.c.}$ respectively. Such terms describe scattering processes between two photons and two or three phonons, and they play an active role in the dynamics only if the frequency of the wall fulfills resonant conditions of the form $2\Omega=\omega_{\bold n}+\omega_{\bold m}$, $2\Omega=\omega_{\bold n}-\omega_{\bold m}$, $3\Omega=\omega_{\bold n}+\omega_{\bold m}$ or $3\Omega=\omega_{\bold n}-\omega_{\bold m}$. 
    \item \textit{Counterrotating terms}. The contributions called counterrottaing terms are $\hat a_{\bold n}\hat a_{\bold m}\hat b^2+\textrm{h.c.}$ and $\hat a_{\bold n}\hat a_{\bold m}\hat b^3+\textrm{h.c.}$ As already noted for the counterrotating terms in $\hat H_1$ \cite{lawInteractionMovingMirror1995,PhysRevA.106.033502}, the contribution of such terms concerns only high-order interactions. In the current scenario, counterrotating terms play a role in processes where the effective Hamiltonian is at least proportional to $\sim \epsilon^3$.
    \item \textit{Cross-Kerr frequency shifts.} Contrary to the terms $\hat H_1$ and $\hat H_3$, the term $\hat H_2$ encodes static contributions, namely terms that are never time-dependent in the interaction picture since they commute with the free Hamiltonian. These terms are the cross-Kerr terms $\hat a_{\bold n}^\dag\hat a_{\bold n}\hat b^\dag\hat b$, and they represent the frequency shift of both the optical and the mechenical modes due to the presence of phonons and photons respectively \cite{chakrabortyEnhancingQuantumCorrelations2017,PhysRevA.91.043822,PhysRevA.99.043837}. Note that in this class of operators we can also include Lamb-shift-like contributions caused by the vacuum energy of both the quantum field and the mechanical oscillator.
    \item \textit{Kerr-like interactions.} The Hamiltonian $\hat H_3$ describes the interaction of an odd number of bosons in a similar fashion to $\hat H_1$. For this reason, we cannot expect static terms (that commute with the free Hamiltonian) in the interaction picture. Nevertheless, we can isolate contributions that display Kerr-like interactions where the interaction itself is mediated by the presence of phonons or photons. Examples of off-resonant terms of this type are $\hat a_{\bold n}^\dag\hat a_{\bold n}\hat b^\dag\hat b(\hat b+\hat b^\dag)$, where the radiation pressure is determined by both the number of photons and phonons. Furthermore, one can also find resonant terms, i.e., those where resonant conditions can lead to time independent coefficients, and they read  $\hat b^\dag\hat b\hat a_{\bold n}^\dag\hat a_{\bold m}^\dag\hat b+\textrm{h.c.}$ or $\hat b^\dag\hat b\hat a_{\bold n}^\dag\hat a_{\bold m}\hat b+\textrm{h.c.}$. Note that resonance conditions required to ``activate'' these terms are the same as those predicted for the interaction Hamiltonian $\hat H_1$.
\end{itemize} 

Finally, the 3-dimensional model discussed above may be reduced to a 1-dimenional model following a simple procedure. In the limit $k_{n_\perp}\rightarrow 0$, the Hamiltonian has again the same formal expression \eqref{quantum:hamiltonian:explicit},
where now each term reads
\begin{align}\label{quantum:hamiltonian:1D}
\hat{H}_0:=&\sum_{n}\,\omega_n\,\hat{a}_n^\dag\hat{a}_n+\Omega\,\hat{b}^\dag\hat{b}\nonumber,\\
\hat{H}_{1}:=&-4\sum_{n,m}(-1)^{n+m}\sqrt{\omega_n\omega_m}\,\hat X_n\hat X_{m}\hat X_{b},\nonumber\\
\hat{H}_{2}:=&8\sum_{n,m}(-1)^{n+m}\sqrt{\omega_n\omega_m}\,\hat X_n\hat X_{m}\hat X_{b}^2,\nonumber\\
\hat{H}_{3}:=&16\sum_{n,m}(-1)^{n+m}\sqrt{\omega_n\omega_m}\left[\frac{k_n k_m L^2}{3}\hat P_n\hat P_{m}\right.\nonumber\\
&\left.+\left(\frac{(k_n^2+k_m^2)L^2}{6}-1\right)\hat X_n\hat X_{m}\right]\hat X_b^3.
\end{align}
In the equations above, we defined $L_\text{x}\equiv L$ and $\omega_n=c k_n\equiv n \pi c/L$. We note that $\hat{H}_{2}$ has recently been obtained for a 1-dimensional cavity \cite{PhysRevA.111.043524}.

\subsection{Perturbed ground state}
In this section we study the ground state of our perturbed Hamiltonian \eqref{quantum:hamiltonian:explicit}. By means of the modified perturbation theory described in Appendix~\ref{perthe}, we are able to calculate the correction to the unperturbed vacuum state at the first order, as well as the correction to the vacuum energy at the second order.

First, we present the expression of the correction $\lvert 0^{(1)}\rangle$ to the vacuum state, which reads
\begin{align}
\lvert 0^{(1)}\rangle=&\sum_{\bold n}\frac{k_{n_\textrm{x}}^2-k_{n_\perp}^2}{2\omega_{\bold{n}}\Omega}\lvert 0_{\bold{n}}0_{\bold{m}} 1_\textrm{b}\rangle\nonumber\\
&+\sum_{\bold n}\frac{\sqrt{2} k_{n_\textrm{x}}^2}{2\omega_{\bold{n}}(2\omega_{\bold{n}}+\Omega)}\lvert 2_{\bold{n}}0_{\bold{m}} 1_\textrm{b}\rangle\nonumber\\
&+\sum_{\bold n\neq\bold m}\frac{(-1)^{n_\textrm{x}+m_\textrm{x}}k_{{n_\textrm{x}}}k_{{m_\textrm{x}}}}{\sqrt{\omega_{\bold n}\,\omega_{\bold m}}(\omega_{\bold{n}}+\omega_{\bold{m}}+\Omega)}\lvert 1_{\bold{n}}1_{\bold{m}} 1_\textrm{b}\rangle.
\end{align}
As expected, the interaction between the cavity field and the wall alters the ground state of the entire system. In particular, we observe that the ground state does not correspond to the vacuum state anymore, but it is a new state populated by virtual particles \cite{PhysRevLett.111.060403}. Note that these particles stem from different terms in the Hamiltonian. The presence of the first term is caused by the radiation pressure of the photonic vacuum state on the wall. This term would disappear if we imposed the normal ordering of the operators of the Hamiltonian $\hat H_1$. The contributions in the last two lines stem from the counterrotating terms of the Hamiltonian. These are normally neglected when considering interactions in the weak-strong coupling, since their presence become relevant only in the strong coupling regime \cite{HABARRIH2023170719,stannigelOptomechanicalQuantumInformation2012, PhysRevResearch.5.013075}.

The energy of the vacuum has perturbative expansion $E_0\simeq E_0^{(0)}+\epsilon^2 E_0^{(2)}$, where we compute $E_0^{(2)}$ that reads 
\begin{align}
E_0^{(2)}=&\sum_{\bold n}\frac{k_{{n_\textrm{x}}}^2}{4\omega_{\bold n}^3}\left(\frac{\zeta_{\bold n \bold n}}{\omega_{\bold n}^2}-\frac{k_{{n_\perp}}^2}{2}\right)-\sum_{\bold n}\frac{\left(k_{n_\textrm{x}}^2-k_{n_\perp}^2\right)^2}{4\omega_{\bold{n}}^2\Omega}\nonumber\\
&-\sum_{\bold n,\bold m}\frac{k_{{n_\textrm{x}}}^2k_{{m_\textrm{x}}}^2}{4\omega_{\bold n}\omega_{\bold m}(\omega_{\bold{n}}+\omega_{\bold{m}}+\Omega)}.
\end{align}
In order to better understand this result, we make a concrete comparison with what has already been achieved in the literature \cite{PhysRevLett.111.060403}. In particular, we reduce the formula above to the case of 1-dimensional cavity, for which $k_{n_\perp}\rightarrow 0$, thereby obtaining 
\begin{align}
E_0^{(2)}=&\sum_{n}\frac{\omega_n}{2}-\sum_{n,m}\frac{\omega_n\omega_m}{4(\omega_n+\omega_m+\Omega)}-\sum_{n}\frac{\omega_{n}^2}{4\Omega}.
\label{encorr}
\end{align}
The last term in the equation above is the energetic contribution of the radiation pressure stemming from the vacuum energy of the field inside the cavity. The second contribution is due to the counterrotating terms of the Hamiltonian, and it encodes the presence of virtual particles that further shift the Casimir energy \cite{PhysRevLett.111.060403}.
Finally, the first term in \eqref{encorr} describes the cross-Kerr energy shift due to the vacuum energy of the field, reciprocally influencing the zero-point energy of the mechanical mode. This contribution stems from the average value of the Hamiltonian $\hat H_2$.

\begin{figure*}[ht!]
	\centering
	\includegraphics[width=0.7\linewidth, angle=270]{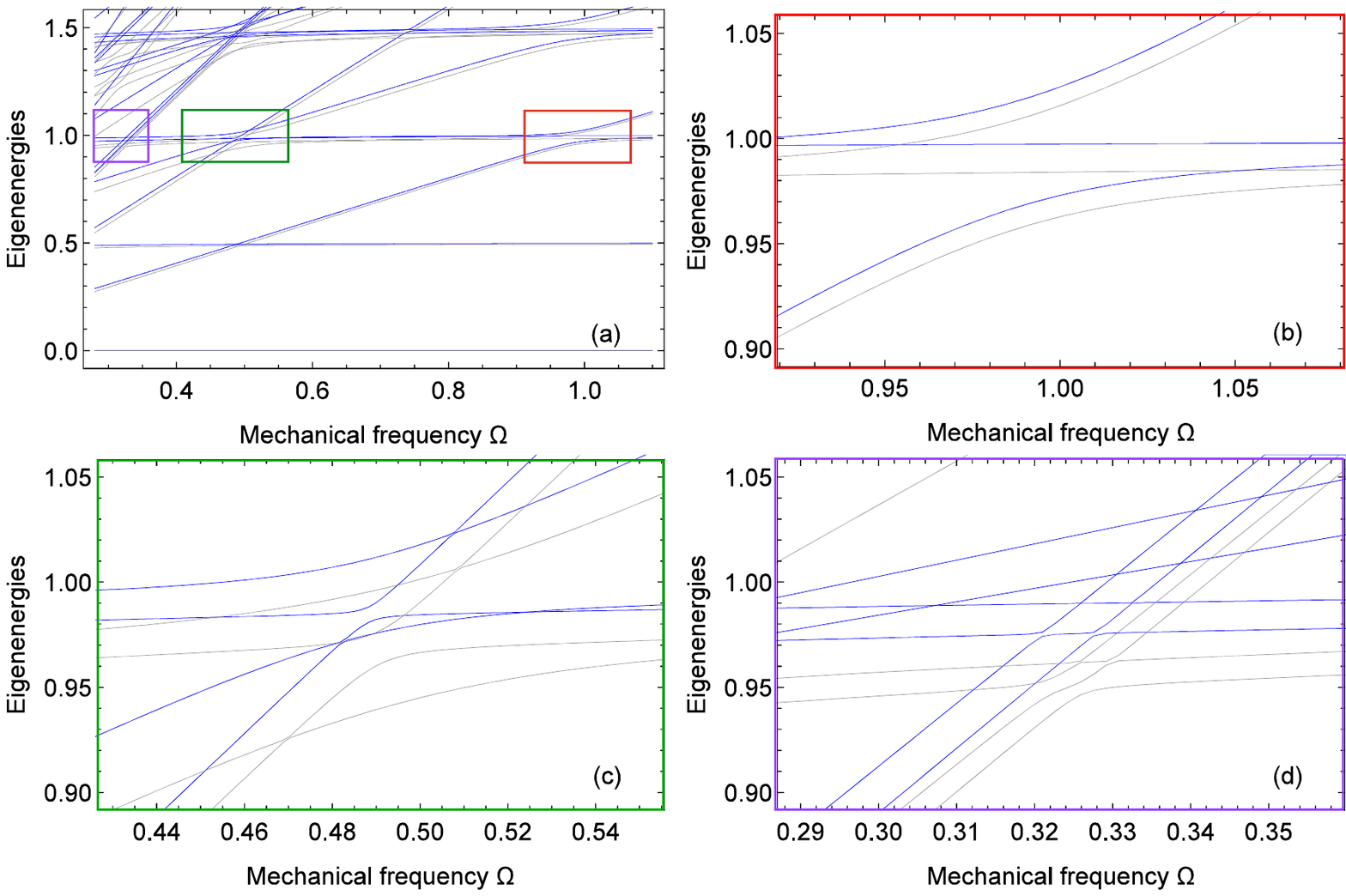}
	\caption{Eigenenergies of the system as a function of the mechanical frequency (a).
    The blue lines represent the eigenvalues of the whole Hamiltonian in \eqref{wholeH}, whereas the gray lines indicate the eigenvalues calculated from \eqref{firstord}. Zoom around the first-order resonance $\Omega=2\omega_1$ (b), the second-order resonance $2\Omega=2\omega_1$ (c), and the third-order resonance $3\Omega=2\omega_1$ (d), respectively. Graphs are realized by setting $\epsilon=0.07$.}
	\label{eigenenergies}
\end{figure*}




\section{Numerical analysis}\label{NA}
The main goal of this section is to examine the effects of high-order interactions encoded in the interaction Hamiltonians $\hat H_2$ and $\hat H_3$ on both the dynamics and the thermodynamics of the system.
To do so, henceforth we restrict our analysis to the three-body interactions between the wall and two cavity modes, $\omega_1$ and $\omega_2$, focusing on the resonant interactions between the mechanical mode and the cavity mode $\omega_1$. Although in a previous work  we have already demonstrated that the presence of the off-resonant mode $\omega_2$ do not contribute significantly to the dynamics \cite{PhysRevResearch.6.023320}, we cannot exclude a priori that the presence of the high-order terms $\hat H_2$ and $\hat H_3$ may alter our previous results.
\subsection{Three-body interaction Hamiltonian}
The Hamiltonian of the optomechanical system describing the interaction of two cavity modes and the quantum mechanical degree of freedom of the wall is $\hat{H}=\hat H_0+\epsilon\hat H_{1}+\epsilon^2\hat H_{2}+\epsilon^3\hat H_{3}$, where 
{\small
\begin{subequations}
\begin{align}
\hat H_0:=&\,\omega_1\hat a_1^\dag\hat a_1+\omega_2\,\hat a_2^\dag\hat a_2+\Omega\,\hat b^\dag\hat b\\
\hat{H}_{1}:=&-4 \,\hat X_{\textrm{b}}\left(\omega_1\,\hat X_1^2+\omega_2\,\hat X_2^2-\sqrt{\omega_1\omega_2}\,\hat X_1\hat X_{2}\right),\label{firstord}\\
\hat{H}_{2}:=&8 \,\hat  X_{\textrm{b}}^2\left(\omega_1\,\hat X_1^2+\omega_2\,\hat X_2^2-\sqrt{\omega_1\omega_2}\,\hat X_1\hat X_{2}\right),\\
\hat{H}_{3}:=&\frac{(4\pi)^2 }{3}\left(\omega_1\hat P_1^2+4\omega_2\hat P_2^2-2\sqrt{\omega_1\omega_2}\hat P_1\hat P_2\right)\hat  X_{\textrm{b}}^3\nonumber\\
&+\frac{(4\pi)^2 }{6}\left(\omega_1\hat X_1^2+8\omega_2\hat X_2^2-5\sqrt{\omega_1\omega_2}\hat X_1\hat X_2\right)\hat X_{\textrm{b}}^3\nonumber\\
&-16 \left(\omega_1\hat X_1^2-\omega_2\hat X_2^2+\sqrt{\omega_1\omega_2}\hat X_1\hat X_2\right)\hat X_{\textrm{b}}^3
\end{align}
\label{wholeH}
\end{subequations}
}
This Hamiltonian can be diagonalized numerically. In particular, in Fig.\ref{eigenenergies}a we plot the eigenenergies calculated with respect to both the total Hamiltonian \eqref{wholeH} (blue lines) and the only first-order contribution \eqref{firstord} (gray lines) as a function of the mechanical frequency $\Omega$, fixing the coupling at $\epsilon=0.07$. To make this plot, we subtracted the energy of the ground state. It is interesting to observe that the spectrum of \eqref{firstord} displays splits of the eigenenergies even when the mechanical frequency is smaller than the optical frequency. Although no terms of the Hamiltonian in \eqref{firstord} suggest explicit resonant interactions when $\Omega\leq\omega_1$, these energy splits are due to high-order interaction processes between the wall and the cavity mode. Such processes are more effective in the strong coupling regime, where the dynamics becomes sensitive to the exchanges of virtual particles \cite{settineriConversionMechanicalNoise2019}.

To analyse such resonant processes in more details, we distinguish resonant interactions according to the order of the coupling parameter $\epsilon$. In particular, henceforth we will refer to:
\begin{itemize}
    \item $\Omega=2\omega_{1}$ as first-order resonance (see Fig.\ref{eigenenergies}b);
    \item $2\Omega=2\omega_{1}$ as second-order resonance (see Fig.\ref{eigenenergies}c);
    \item $3\Omega=2\omega_{1}$ as third-order resonance (see Fig.\ref{eigenenergies}d).
\end{itemize}
In the second-order (third-order) resonance the factor 2 (3) in front of $\Omega$ highlights the fact that the creation of the photon pair occurs at the price of two (three) phonons.

The presence of $\hat H_2$ and $\hat H_3$ in the Hamiltonian has a clear impact on the energy levels of our quantum system. In particular, we distinguish two main effects: the shift of the energy level, and the modification of the resonant interactions. The first effect is evident from the fact that blue and gray lines in Fig.\ref{eigenenergies} never overlap. Such energy shift is due to Kerr-shifts caused by the presence of the static contributions in $\hat H_2$. 
To observe the modifications of the resonant interactions, we need to focus our attention on the energy splits. Interestingly, whereas the first-order interaction seems to not be influenced by the presence of $\hat H_2$ and $\hat H_3$ (see Fig.\ref{eigenenergies}b), the energy splits in Fig.\ref{eigenenergies}c and Fig.\ref{eigenenergies}d are dramatically altered. 

To better understand the different impact of $\hat H_1$, $\hat H_2$ and $\hat H_3$ on the energy spectrum, we recall that, in case of the first-order resonant interaction $\Omega=2\omega_1$, the dominant contribution to the dynamics (and in particular to the boson scattering) is given by those terms of the Hamiltonian that do not rotate in the interaction picture, namely those proportional to $\hat b(\hat a_1^\dag)^2+\hat b^\dag \hat a_1^2$. Such Casimir-like terms permit the conversion of single phonons into real photons, and are already included in $\hat H_1$, namely the interaction Hamiltonian at the lowest order in $\epsilon$. Any alteration of the spectrum due to the presence of $\hat H_2$ and $\hat H_3$ (apart from the Kerr-shift) is negligible. 
On the contrary, high-order resonant interactions are influenced by all the interaction Hamiltonians $\hat H_1$, $\hat H_2$ and $\hat H_3$. Indeed, fixed $2\Omega=2\omega_1$ or $3\Omega=2\omega_1$, the photon-phonon conversion takes place via both the creation and annihilation of virtual (off-shell) particles predicted in high-order perturbative processes predicted by $\hat H_1$, and the conversion of real particles enabled by $\hat H_2$ and $\hat H_3$.

The competition of virtual and real processes in high-order resonant interactions has an impact on the effective coupling strength. To quantify it, we can calculate the effective Hamiltonian by means of  James' method \cite{shaoGeneralizedJamesEffective2017}. Starting from the Hamiltonian in \eqref{wholeH} and focusing only on the second-order resonance $2\Omega=2\omega_1$, it can be seen that the effective interaction Hamiltonian is $\hat H_{\textrm{eff}}=\epsilon^2\hat H_{\textrm{Kerr}}+\epsilon^2\hat H_{\textrm{scat}}$, where $\hat H_{\textrm{Kerr}}$ includes all Kerr-like frequency shifts, whereas 
\begin{align}
\hat H_{\textrm{scat}}=\frac{\omega_1}{2}\left[\frac{2}{\Omega}\left(\omega_1+\frac{\omega_2}{8}\right)+1\right]\left[\hat a_1^2\hat b^\dag{}^2+\hat a_1^\dag{}^2\hat b^2\right],
\end{align}
represents the two-phonon-two-photon scattering.
The first term of the coupling strength, namely that proportional to $2/\Omega$, stems from virtual processes mathematically described in the interaction picture by the effective Hamiltonian $-i\hat H_1(t)\int \hat H_1(t')dt'$ \cite{shaoGeneralizedJamesEffective2017}. The second term is already contained in $\hat H_2$. It is easy to observe that, once the resonance conditions $\omega_2=2\omega_1=2\Omega$ are imposed, the coupling constant is determined at 71.4\% by virtual processes arising from $\hat H_1$, and at 28.6\% by $\hat H_2$. This implies that more than 1/4 of the coupling strength is given by $\hat H_2$.
\subsection{Dynamics and thermodynamics}
In this section, we investigate the heat transfer between the hot bath coupled to the wall and the cold bath coupled to the cavity by analysing the dynamics of the populations of the involved bosonic degrees of freedom, the rates of heat transferred between modes and baths, as well as the entropy production rate.
To do so, we study the time-evolution of the system in two different scenarios: in one case, we take into account the whole Hamiltonian in \eqref{wholeH}; in the other, we consider the evolution solely due to \eqref{firstord}. According to our analysis of the energy spectra, we should expect that the presence of $\hat H_2$ and $\hat H_3$ in the Hamiltonian has a greater impact on the dynamics ruled by high-order interactions than on the dynamics first-order one.

Results have been obtained by first numerically solving the master equation in \eqref{me} appling numerical filtering technique with parameter $\Delta=0.09$ normalized with respect to $\omega_2$ (details on the master equation are discussed in Appendix \ref{master}), and subsequently calculating the average values of each quantity of interest. In particular, thermodynamic quantities such as heat flows and entropy production rate have been defined in Appendix \ref{thermo}, respectively in \eqref{hf} and \eqref{epr}. For the realization of all graphs we have used the following parameters:
$\omega_{2}=2\omega_{1}=1$, $\gamma=0.018$, $\kappa=0.006$, $T_{\textrm{c}}=10^{-6}$ and $T_{\textrm{w}}=0.3$. Moreover, for mere convenience matter, we have chosen $\kappa_0=0.003$ as time scale. The quantum mechanical frequency of the wall is determined by the resonant conditions, and it is therefore specified on the caption of each graph. Finally, frequencies and temperatures have been normalized with respect to $\omega_{2}$.
\begin{figure}[ht!]
	\centering
	\includegraphics[width=1\linewidth]{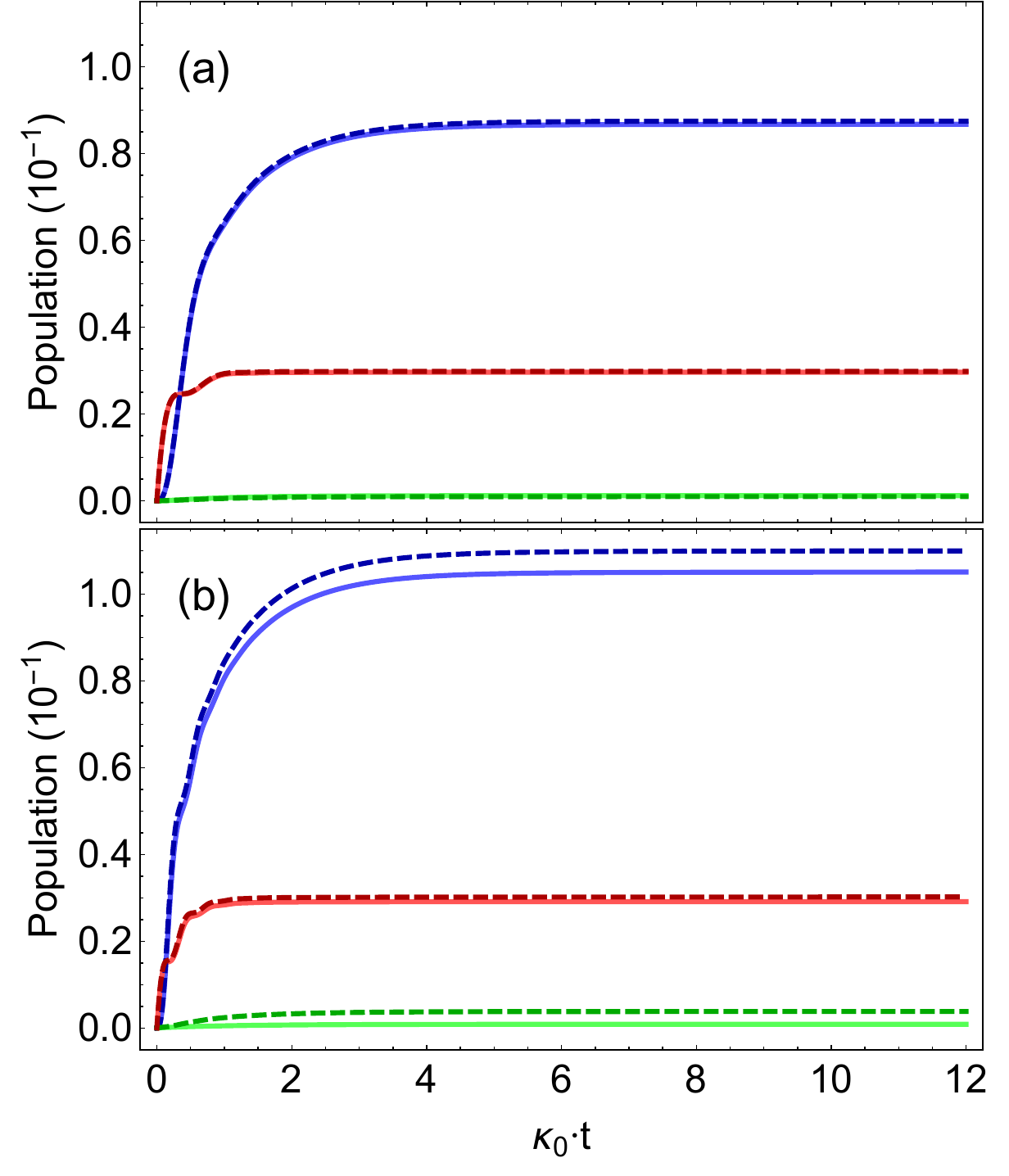}
	\caption{Time evolution of the populations of the mirror (red line), of cavity mode 1 (blue line), and of cavity mode 2 (green line), fixing the resonant condition $\Omega=2\omega_1$. 
    The solid lines represent the dynamics of the populations evolving via the whole Hamiltonian in \eqref{wholeH}, whereas the dashed lines indicate the evolution due to the only contribution in \eqref{firstord}. The two graphs are realized by setting $\epsilon=0.03$ (a), and $\epsilon=0.07$ (b).}
	\label{grvseqp}
\end{figure}
\begin{figure}[ht!]
	\centering	\includegraphics[width=1\linewidth]{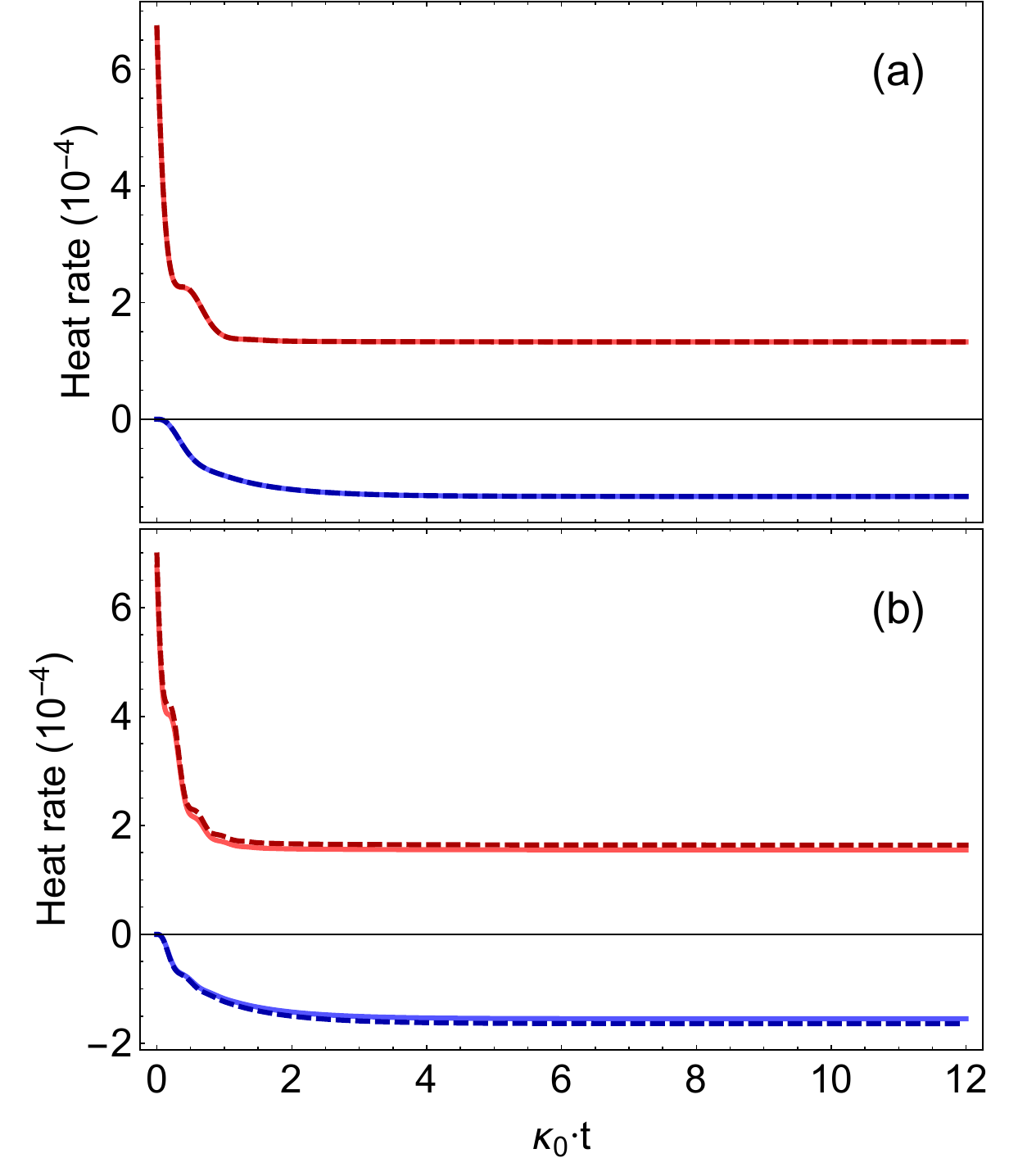}
	\caption{Time evolution of $\mathcal{J}_{\text{w}}(t)$ (red line) and $\mathcal{J}_{\text{c}}(t)$ (blue line) after enforcing the resonant condition $\Omega=2\omega_1$. 
    The solid lines represent the dynamics of the heat flows evolving via the whole Hamiltonian in \eqref{wholeH}, whereas the dashed lines indicate the evolution due to the only contribution in \eqref{firstord}. The graphs are realized by setting $\epsilon=0.03$ (a), and $\epsilon=0.07$ (b).}
	\label{grvseqh}
\end{figure}
\begin{figure}[ht!]
	\centering
\includegraphics[width=1\linewidth]{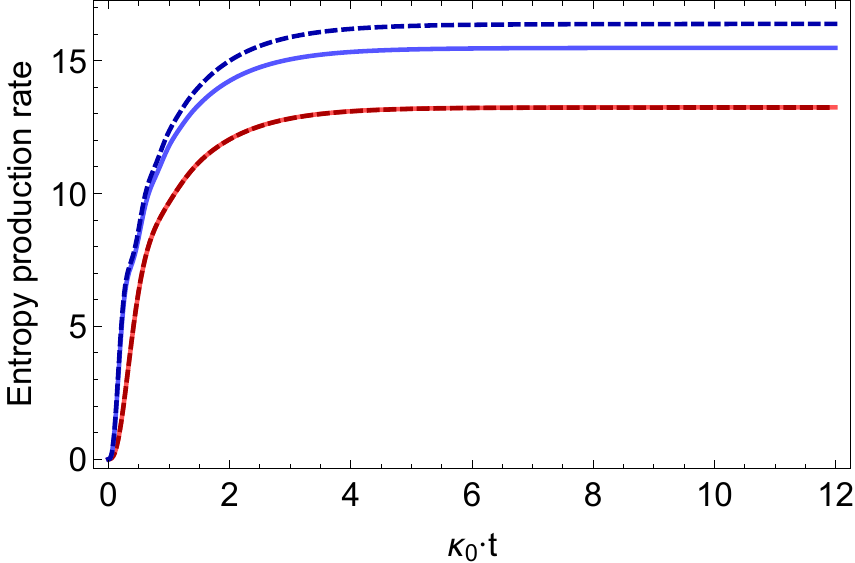}
	\caption{Time evolution of the entropy production rate fixing the resonant condition $\Omega=2\Omega_1$.
    The solid lines represent the dynamics of the entropy production rate evolving via the whole Hamiltonian in \eqref{wholeH}, whereas the dashed lines indicate the evolution due to the only contribution in \eqref{firstord}. Plots have been realized by setting $\epsilon=0.03$ (red lines), and $\epsilon=0.07$ (blue lines).}
	\label{grvseqe}
\end{figure}
\subsubsection{First-order resonance}
We start from the time-evolution of populations, heat rates and the entropy production rate, which are plotted in Fig.\ref{grvseqp}, Fig.\ref{grvseqh} and Fig.\ref{grvseqe} respectively, when the first-order resonant condition is fulfilled. In particular, the first aspect we want to discuss is the phononic population.

Although we cannot make a precise analytical prediction of the phonon number in the steady state, we can still provide a reasonable estimation. In the strong coupling regime, the coupling strength between the two subsystems is much lower than the bare frequencies of the involved bosonic modes, but it is higher than any loss rates. Since we assumed $\kappa<\gamma$, the wall tends to thermalize to its own bath at $T_\textrm{w}$, however, the interaction constantly forces a relatively small part of the phononic population to convert into photons. The phononic population is therefore slightly smaller than the average value expected by the Bose-Einstein distribution for a single, non-interacting boson. The decrease of the phononic population has already been discussed for the first-order resonance in the literature \cite{PhysRevResearch.6.023320}, where it has been proposed as a form of cooling mechanism of the mechanical mode. Note that the bath at $T_\textrm{c}\simeq 0$ and the interaction with the wall induce the number of photon to depend on the effective coupling strength and the loss rates.

The coupling strength clearly plays a crucial role in the dynamics.
It has already been shown that, in the strong coupling regime, the coupling strength $\epsilon$ affects both the photonic population and the heat flows between the two subsystems at the steady state \cite{PhysRevResearch.6.023320}. This is observed again here: the expected population of the optical modes when $\epsilon=0.07$ (see Fig.\ref{grvseqp}b) is higher than expected population when $\epsilon=0.03$ (see Fig.\ref{grvseqp}a). Moreover, the heat rate $\mathcal{J}_{\text{w}}(t)$ (and consequently also $\lvert \mathcal{J}_{\text{c}}(t)\rvert$) in Fig.\ref{grvseqh}b is slightly higher than in Fig.\ref{grvseqh}a, suggesting that stronger interactions between the two subsystems lead to the enhancement of the heat flows between the two baths. This becomes more evident when we look at the entropy production rates in Fig.\ref{grvseqe}: once the steady state has been reached, the blue curves ($\epsilon=0.07$) are drastically higher than the red curves ($\epsilon=0.03$). Recalling that the two subsystems are coupled to two baths at different temperatures, we have that the stronger the wall-field interactions, the more efficient the heat propagation. This suggests that the entropy production rate not only determines the heat-transfer efficiency, but it can represent a valid instrument for the estimation of the effective coupling between the subsystems.

Another fundamental aspect to consider is the weight of both $\hat H_2$ and $\hat H_3$ in the dynamics. We see that the influence of such high-order terms to the dynamics strongly depends on the intensity of $\epsilon$. Whereas dashed and solid curves in Fig.\ref{grvseqp}a and Fig.\ref{grvseqh}a, namely when the coupling is $\epsilon=0.03$, perfectly overlap, they split in Fig.\ref{grvseqp}b and \ref{grvseqh}b, namely when the coupling is $\epsilon=0.07$. This indicates that the presence of $\hat H_2$ and $\hat H_3$ affects the dynamics at higher coupling, and that they can be safely ignored in proximity of the weak coupling regime, namely when $\epsilon\,\omega_1\sim \gamma$. Finally, although dashed and solid heat rate curves fundamentally overlap in Fig.\ref{grvseqh}b, we can again extract more information from the entropy production rate. In particular, we observe that the blue dashed curve in Fig.\ref{grvseqe} reaches a higher value at the steady state than the solid blue curve. This suggests that, within the conditions established by the first-order resonance, the first-order interaction Hamiltonian $\hat H_1$ slightly overestimates the real interactions between the two subsystems.

\begin{figure}[ht!]
	\centering
	\includegraphics[width=1\linewidth]{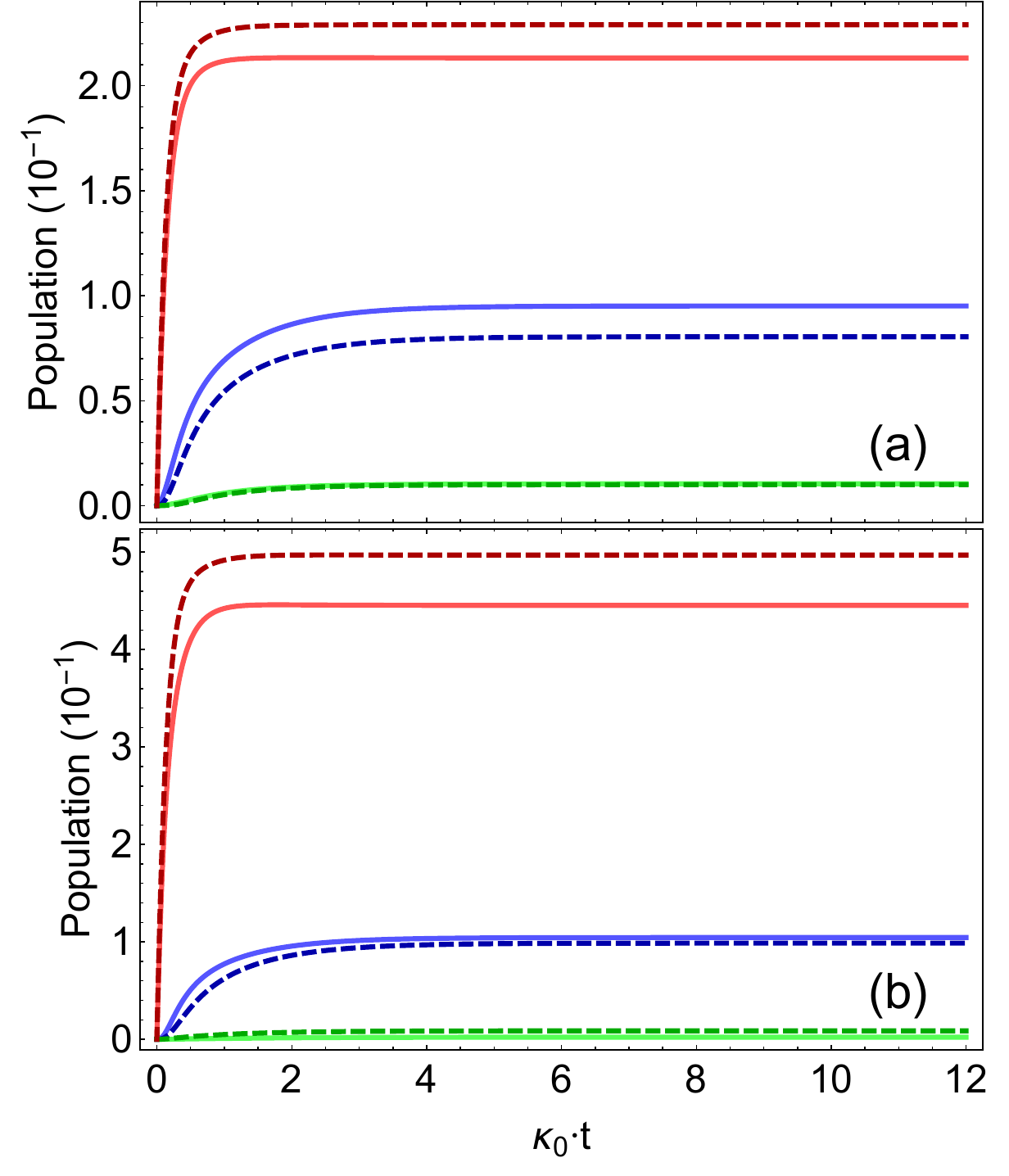}
	\caption{Time evolution of the populations of the mirror (red line), of the cavity mode 1 (blue line), and of cavity mode 2 (green line), fixing the resonant conditions $\Omega=\omega_{1}$ (a) and $3\Omega=2\omega_{1}$ (b). 
    The solid lines represent the dynamics of the populations evolving via the whole Hamiltonian in \eqref{wholeH}, whereas the dashed lines indicate the evolution due to the only contribution in \eqref{firstord}.}
	\label{p22and32}
\end{figure}

\subsubsection{Second- and third-order resonances}
While the presence of $\hat H_2$ and $\hat H_3$ scarcely affects the outcomes expected when imposing the first-order resonance, results drastically change once we activate second- and third-order resonance conditions. 

First, we notice the dramatic change in the phononic population in Fig.\ref{p22and32} with respect to Fig. \ref{grvseqe}b. This does not come as a surprise: in order to activate second- and third-order resonances we needed to adjust the mechanical frequency accordingly. This inevitably leads to a drastic change in the phonon number in accordance to the Bose-Einstein distribution.
\begin{figure}[ht!]
	\centering
\includegraphics[width=1\linewidth]{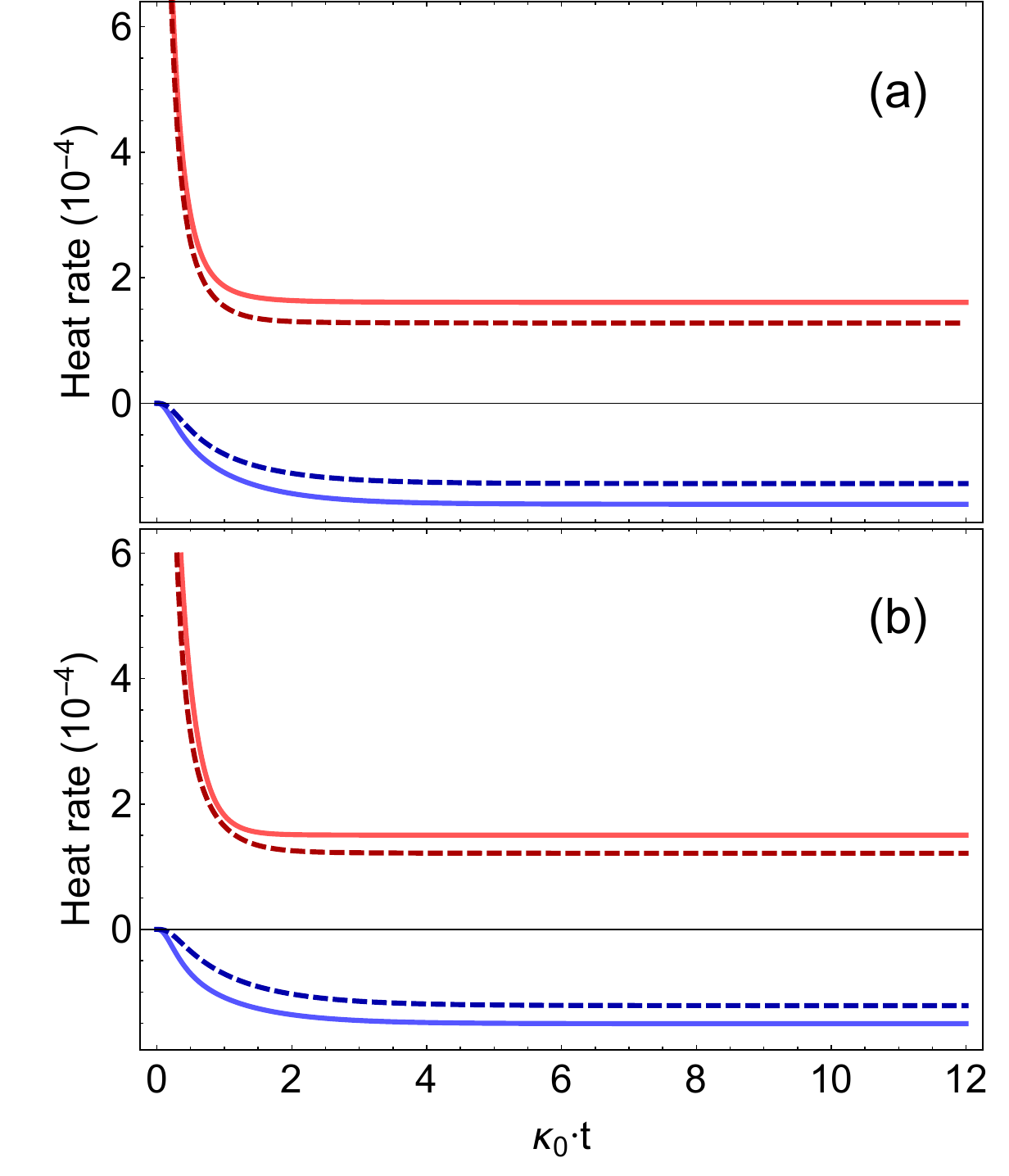}
	\caption{Time evolution of $\mathcal{J}_{\text{w}}(t)$ (red line) and $\mathcal{J}_{\text{c}}(t)$ (blue line) fixing the resonant conditions $\Omega=\omega_{1}$ (a) and $3\Omega=2\omega_{1}$ (b). 
    The solid lines represent the dynamics of the heat flows evolving via the whole Hamiltonian in \eqref{wholeH}, whereas the dashed lines indicate the evolution due to the only contribution in \eqref{firstord}.}
	\label{h22and32}
\end{figure}

\begin{figure}[ht!]
	\centering
	\includegraphics[width=1\linewidth]{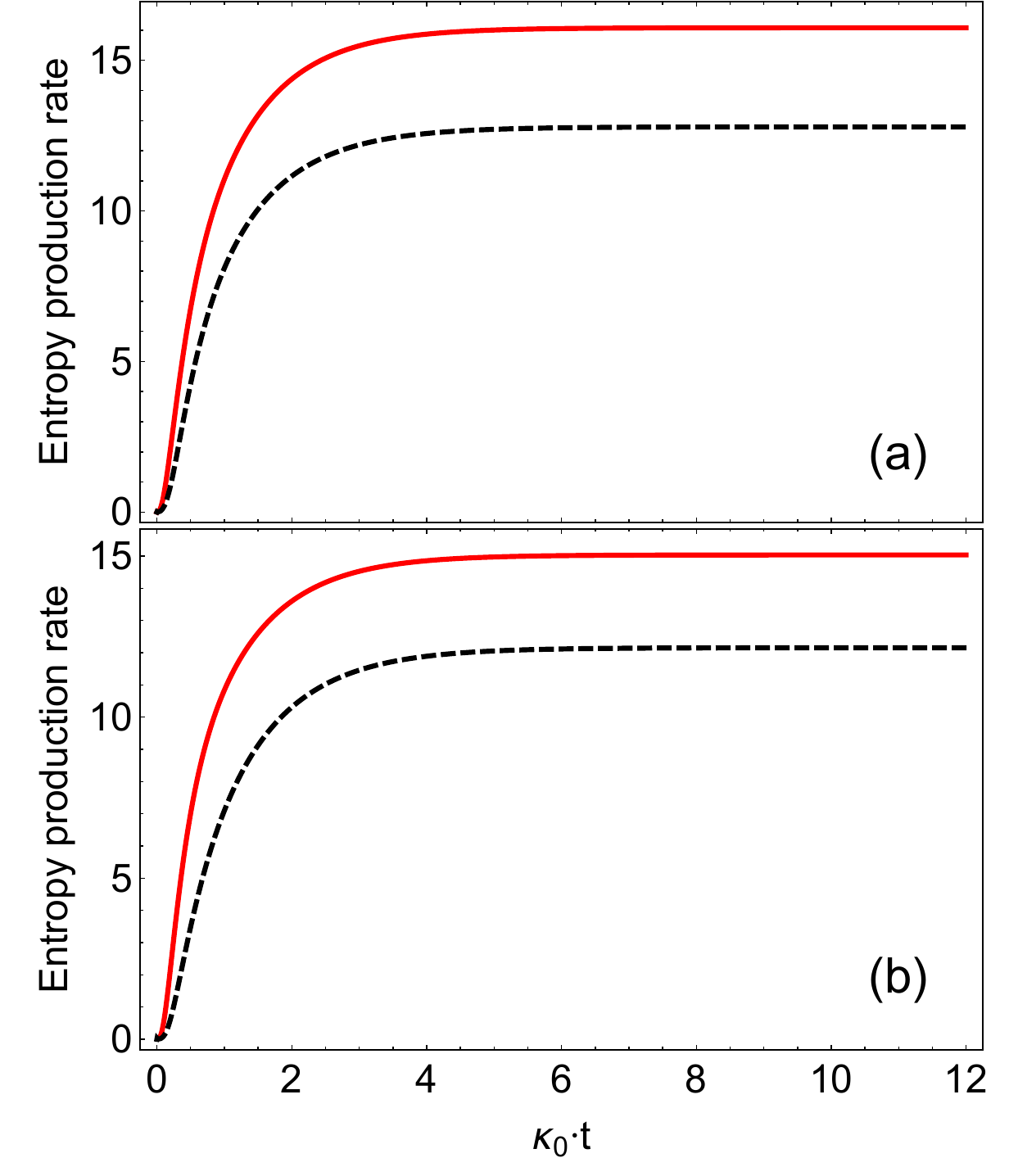}
	\caption{Time evolution of the entropy production rate, fixing the resonant conditions $\Omega=\omega_{1}$ (a) and $3\Omega=2\omega_{1}$ (b).
    The red solid line represents the dynamics of the entropy production rates evolving via the whole Hamiltonian in \eqref{wholeH}, whereas the black dashed line indicates the evolution due to the only contribution in \eqref{firstord}.}
	\label{e22and32}
\end{figure}

More than the phononic population itself, the relevant aspect we want to discuss here is the sensitivity of the phononic population to the high-order Hamiltonian terms. In the previous section we have observed that, once we have fixed the first-order resonance and $\epsilon=0.07$, the presence of $\hat H_2$ and $\hat H_3$ (slightly) changes the photonic population, while the phononic population is scarcely affected. Instead, we now observe that the number of phonons changes notably, which seems to be a symptom of a deep modification of the effective coupling between mechanical and optical subsystems. Note that the change of coupling strength is witnessed also by the heat rates and the entropy production rates, plotted in Fig.\ref{h22and32} and Fig.\ref{e22and32} respectively. Moreover, the dependence of both the number of phonons and photons on the effective coupling, once  the resonance condition $2\Omega=2\omega_1$ has been imposed, has already been discussed in the literature \cite{settineriConversionMechanicalNoise2019}, where it has also been observed that the increase of the coupling strength induces a higher conversion of hot phonons into photons.

However, the sole variation of the effective coupling strength is not sufficient to justify the drastic dependence of the phononic population on $\hat H_2$ and $\hat H_3$. Indeed, if this were the case, we should have observed a similar behaviour in Fig.\ref{grvseqp} and in \cite{PhysRevResearch.6.023320}, where on the contrary the phononic population seems to be unaffected by the coupling strength. A more intuitive explanation is that, in the first-order resonance, the wall constantly absorbs single hot phonons from the bath at the rate dictated by $\gamma$, and converts single phonons into photons. In this way, the two processes, namely the phonon absorption from the bath and the phonon-photon conversion, are balanced, and the phononic population is only impacted by the mechanical frequency $\Omega$ and the coupling to the baths.
On the contrary, in high-order resonances the phonon-photon conversion involves the scattering of two or three phonons at the same time, while the phonon absorption from the hot bath occurs at the same rate. The conversion of a higher number of phonons makes therefore the phononic population more sensitive to the effective coupling.

It is interesting to notice that such higher sensitivity of the phononic population to the coupling strength observed in high-resonant interactions makes the cooling mechanism proposed in the literature more appealing, see \cite{PhysRevResearch.6.023320} . Indeed, the use of high-order resonances makes the number of photons susceptible to the effective wall-cavity coupling, thereby providing an additional parameter by means of which we can control the cooling effect. Moreover, high-order resonances require lower mechanical frequencies, which is of great advantage for future experimental implementations.

\section{Conclusions}
High-order resonant interactions mediated by virtual processes in optomechanical systems are becoming of increasing interest  \cite{stannigelOptomechanicalQuantumInformation2012,distefanoInteractionMechanicalOscillators2019,russoOptomechanicalTwophotonHopping2023,10.21468/SciPostPhys.18.2.067}. Nevertheless, to date such high-order interactions between mirror and field have always been deducted starting from Hamiltonian models that encode linear wall-field interactions in $\epsilon$.

The linearity of the coupling with respect to $\epsilon$ has always been a valid approximation of the optomechanical setups that model accessible experimental regimes, namely as long as both the mechanical frequency and the coupling strength are much lower than any optical frequencies. 
However, these conditions imply that, on the one hand, the dynamics is essentially determined by linear interactions only, while, on the other, high-order (virtual) processes can be safely ignored.
The mathematical formulation of high-order processes is therefore inevitably bound to the issue of the linearity of the optomechanical coupling. 

The main claim of this work is  that the presence of high-order nonlinear interactions do not occur only via virtual processes but also via the direct nonlinear wall-field coupling. We have also shown that the dynamics of the system predict different thermodynamic behaviour when the model includes quadratic and cubic terms in the small relative cavity-length fluctuation $\epsilon$.

The importance of the results discussed in this work does not rely only on the fact that high-order interactions have already been observed experimentally in optomechanical apparatuses \cite{fongPhononHeatTransfer2019}, but also on the possibility to predict similar effects in other quantum systems already characterized by optomechanical-like interactions, such as trapped atoms \cite{izadyariQuantumSignaturesQuadratic2022}, phononic-crystal resonator \cite{spinnler2024single}, and superconducting circuits \cite{johanssonOptomechanicallikeCouplingSuperconducting2014}. We believe that our work opens the way to the realization of strong multipartite entangled states in optomechanical systems and exotic quantum heat engines based on high-order interactions.

\section{Acknowledgment}
A.F. and D.E.B. acknowledge support from the joint project No. 13N15685 ``German Quantum Computer based on Superconducting Qubits (GeQCoS)'' sponsored by the German Federal Ministry of Education and Research (BMBF) under the \href{https://www.quantentechnologien.de/fileadmin/public/Redaktion/Dokumente/PDF/Publikationen/Federal-Government-Framework-Programme-Quantum-technologies-2018-bf-C1.pdf}{framework programme
``Quantum technologies -- from basic research to the market''}. D.E.B. also acknowledges support from the German Federal Ministry of Education and Research via the \href{https://www.quantentechnologien.de/fileadmin/public/Redaktion/Dokumente/PDF/Publikationen/Federal-Government-Framework-Programme-Quantum-technologies-2018-bf-C1.pdf}{framework programme
``Quantum technologies -- from basic research to the market''} under contract number 13N16210 ``SPINNING''.
V.M. acknowledge PNRR MUR project “National Quantum Science and Technology Institute” – NQSTI (Grant No. PE0000023).
\

\bibliography{Alexandria}

\appendix
\onecolumngrid

\section{Theoretical background}
\subsection{Calculation of the Hamiltonian}\label{H3app}
In this section we present the mathematical procedure to obtain the Hamiltonian of the optomechanical system. This technique, which has been inspired by the work of Law \cite{lawInteractionMovingMirror1995}, is valid for any $(1+N)$-dimensional system, where $N$ is the spatial dimension, and it has been studied in more detail for bosonic fields \cite{PhysRevA.106.033502}, as well as fermionic fields \cite{PhysRevA.106.052204}. In the regime of small oscillations such formalism permits the calculation of the interaction Hamiltonian up to any order in the oscillation amplitude of the mirror by means of a perturbative approach without therefore solving any differential equation of motion \cite{PhysRevA.106.033502}.

We consider a massless scalar field confined in a 3-D cavity as a simple model for the electromagnetic field to obtain a qualitative understanding of the dynamics. The Lagrangian of the field is
\begin{align}\label{lagrangian:density}
\mathcal{L}(t,\bold x)=\frac{1}{2}\partial_\mu\phi\partial^\mu\phi.
\end{align}
The field confinement is imposed by means of boundary conditions to the equations of motion. If one of the cavity walls oscillates, say along the direction $x$, the wall+field system is described by coupled differential equations for both the wall and the field. However, if the oscillation amplitude of the wall $\delta L_\text{x}$ is much smaller than the cavity length $L_\text{x}$, i.e., $\delta L_\text{x}/L_\text{x}\ll1$, it results much simpler to first assume that the wall is fixed, thereby solving the equation of motion for the field with static Dirichlet boundary conditions, and only subsequently introduce a small fluctuation in the wave vector of the field modes. 

To solve the equation of motion, namely the Klein Gordon equation $\partial_t^2\phi-\nabla^2\phi$=0, we consider static Dirichlet boundary conditions. The equation of motion admits as a solution
\begin{align}\label{field:expression}
\phi(t,\bold x)=&\sum_{\bold n}\left[\alpha_{\bold n}\,\phi_{\bold n}(t,x)+\alpha_{\bold n}^*\,\phi^*_{\bold n}(t,x)\right],
\end{align}
where each mode $\phi_{\bold n}(t,\bold x)$ is described by
\begin{align}\label{field:modes}
\phi_{\bold n}(t,\bold x)=&\sqrt{\frac{4}{\omega_{\bold n} V}}e^{-i\omega_{\bold n}t}\chi_{\bold n}(\bold x)
\end{align}
with
\begin{align}
\chi_{\bold n}(\bold x)=&\sin\left(\frac{n_x\pi}{L_x}x\right)\sin\left(\frac{n_y\pi}{L_y}y\right)\sin\left(\frac{ n_z\pi}{L_z}z\right)=\sin\left(k_{{n_\textrm{x}}}x\right)\sin\left(k_{{n_\textrm{y}}}y\right)\sin\left(k_{{n_\textrm{z}}}z\right),
\end{align}
and it is characterized by the dispersion relation
\begin{align}\label{field:disp}
\omega_{\bold n}:=\sqrt{k_{{n_\textrm{x}}}^2+k_{{n_\textrm{y}}}^2+k_{{n_\textrm{z}}}^2},
\end{align}
with $\bold n\equiv(n_x,n_y,n_y)$ a set of positive integer numbers and we have introduced the components of the wave vectors. In \eqref{field:modes}, we introduced the volume $V=L_x L_y L_z$. 

We can calculate Hamiltonian density $\mathcal{H}(t,\bold x)=\frac{1}{2}\left[\Pi^2(t,\bold x)+(\nabla\phi(t,\bold x))^2\right]$, with canonical momentum $\Pi(t,\bold x):=-\partial_t\phi(t,\bold x)$.
This reads
\begin{align}
\mathcal{H}(t,\bold x)=&\frac{1}{V}\sum_{\bold n \bold m}\left[-\sqrt{\omega_{\bold n}\omega_{\bold m}}(\alpha_{\bold n}^*-\alpha_{\bold n})(\alpha_{\bold m}^*-\alpha_{\bold m})\sin\left(k_{{n_\textrm{x}}}x\right)\sin\left(k_{{n_\textrm{y}}}y\right)\sin\left(k_{{n_\textrm{z}}}z\right)\sin\left(k_{{m_\textrm{x}}}x\right)\sin\left(k_{{m_\textrm{y}}}y\right)\sin\left(k_{{m_\textrm{z}}}z\right)\right.\nonumber\\
&+\frac{k_{{n_\textrm{x}}}k_{{m_\textrm{x}}}}{\sqrt{\omega_{\bold n}\omega_{\bold m}}}(\alpha_{\bold n}^*+\alpha_{\bold n})(\alpha_{\bold m}^*+\alpha_{\bold m})\cos\left(k_{{n_\textrm{x}}}x\right)\sin\left(k_{{n_\textrm{y}}}y\right)\sin\left(k_{{n_\textrm{z}}}z\right)\cos\left(k_{{m_\textrm{x}}}x\right)\sin\left(k_{{m_\textrm{y}}}y\right)\sin\left(k_{{m_\textrm{z}}}z\right)\nonumber\\
&+\frac{k_{{n_\textrm{y}}}k_{{m_\textrm{y}}}}{\sqrt{\omega_{\bold n}\omega_{\bold m}}}(\alpha_{\bold n}^*+\alpha_{\bold n})(\alpha_{\bold m}^*+\alpha_{\bold m})\sin\left(k_{{n_\textrm{x}}}x\right)\cos\left(k_{{n_\textrm{y}}}y\right)\sin\left(k_{{n_\textrm{z}}}z\right)\sin\left(k_{{m_\textrm{x}}}x\right)\cos\left(k_{{m_\textrm{y}}}y\right)\sin\left(k_{{m_\textrm{z}}}z\right)\nonumber\\
&\left.+\frac{k_{{n_\textrm{z}}}k_{{m_\textrm{z}}}}{\sqrt{\omega_{\bold n}\omega_{\bold m}}}(\alpha_{\bold n}^*+\alpha_{\bold n})(\alpha_{\bold m}^*+\alpha_{\bold m})\sin\left(k_{{n_\textrm{x}}}x\right)\sin\left(k_{{n_\textrm{y}}}y\right)\cos\left(k_{{n_\textrm{z}}}z\right)\sin\left(k_{{m_\textrm{x}}}x\right)\sin\left(k_{{m_\textrm{y}}}y\right)\cos\left(k_{{m_\textrm{z}}}z\right)\right]
\end{align}

Having obtained the Hamiltonian density, the procedure to quantize the optomechanical system in \cite{PhysRevA.106.033502} provides for the following steps:
\begin{enumerate}
    \item Allow for a small fluctuation of the cavity length by replacing $L_\text{x}\rightarrow L_\text{x}+\delta L_\text{x}$, and then expanding each wave vector $k_\text{x}$ in the the Hamiltonian density with respect to the perturbation parameter $\delta L_\text{x}/L_\text{x}$ up to the third order.
    \item Integrate the Hamiltonian density over the box volume.
    \item Promote the field mode amplitudes $a_{\bold n}$ and the small perturbation $\delta L_\text{x}$ to the quantum operators $\hat a_{\bold n}$ and $\delta L_0(\hat b+\hat b^\dag)$, respectively. 
\end{enumerate}
We do not go through the first and the second steps into details, as they are lengthy calculations merely consisting of Maclaurin expansions and integrations over the spatial variables.
The last step is complemented by imposing canonical bosonic commutation rules for the oscillation operators $[\hat b,\hat b^\dag]=1$ as well as for the field modes $[\hat a_{\bold n},\hat a_{\bold m}^\dag]=\delta_{\bold n \bold m}$. Note that $\bold n=(n_\text{x},n_\text{y},n_\text{z})$.
We stress that, in this protocol, we expand the Hamiltonian density with respect to $\delta L_\text{x}$ only in the wave vectors. This is equivalent to the procedure introduced originally by Law \cite{lawInteractionMovingMirror1995}, where both the bosonic operators $\hat a_k$ and the mode frequencies $\omega_k$ have been expanded with respect to the oscillation amplitude $x_m$ around the average cavity length $l_0$.

Once terminated this procedure, we obtain the Hamiltonian in \eqref{quantum:hamiltonian:explicit} and \eqref{quantum:hamiltonian:terms2}. In this expression, we defined three functions of the wave vectors, $\zeta_{\bold n \bold m}$, $\theta_{\bold n \bold m}$ and $\theta_{\bold n \bold m}$. In particular, the interaction sector $\hat{H}_{2}$ contains the function
\begin{align}
\zeta_{\bold n \bold m}=&k_{{m_\textrm{x}}}^2(2k_{{n_\textrm{x}}}^2+3k_{n_\perp}^2)+k_{n_\perp}^2(3k_{{n_\textrm{x}}}^2+4k_{n_\perp}^2),
\end{align}
whereas the interaction Hamiltonian $\hat{H}_{3}$ contains the two functions
\begin{align}
\theta_{\bold n \bold m}=&4 k_{{m_\textrm{x}}}^6 \omega_{\bold n}^4 L_\textrm{x}^2 + k_{{m_\textrm{x}}}^4 \{4 k_{{n_\textrm{x}}}^6 L_\textrm{x}^2 - k_{n_\perp}^4 (45 + 8 \omega_{\bold m} \omega_{\bold n}L_\textrm{x}^2) \nonumber\\
&+ 2 k_{{n_\textrm{x}}}^2 k_{n_\perp}^2 [-27 +2 (k_{n_\perp}^2 -4 \omega_{\bold m} \omega_{\bold n}) L_\textrm{x}^2]+8 k_{{n_\textrm{x}}}^4 [-3 + (k_{n_\perp}^2 -\omega_{\bold m}\omega_{\bold n})L_\textrm{x}^2]\}\nonumber\\
& - k_{n_\perp}^4 \{-4 k_{{n_\textrm{x}}}^6 L_\textrm{x}^2 +k_{{n_\textrm{x}}}^4 (45 +8\omega_{\bold m} \omega_{\bold n}L_\textrm{x}^2) \nonumber\\
&+ 2 k_{{n_\textrm{x}}}^2 k_{n_\perp}^2 [51 + 6 k_{n_\perp}^2 L_\textrm{x}^2 +8 \omega_{\bold m} \omega_{\bold n}L_\textrm{x}^2] + 8 k_{n_\perp}^4 [9 + (k_{n_\perp}^2 +\omega_{\bold m} \omega_{\bold n})L_\textrm{x}^2]\} \nonumber\\
&-2 k_{{m_\textrm{x}}}^2 k_{n_\perp}^2 \{-4 k_{{n_\textrm{x}}}^6 L_\textrm{x}^2 +k_{{n_\textrm{x}}}^4 [27 - 2 k_{n_\perp}^2 L_\textrm{x}^2 + 8 \omega_{\bold m} \omega_{\bold n}L_\textrm{x}^2]\nonumber\\
&+ k_{n_\perp}^4 [51 + 6 k_{n_\perp}^2 L_\textrm{x}^2 +8 \omega_{\bold m} \omega_{\bold n}L_\textrm{x}^2] + k_{{n_\textrm{x}}}^2 (k_{n_\perp}^2) [63 +8 (k_{n_\perp}^2+2 \omega_{\bold m} \omega_{\bold n})L_\textrm{x}^2]\}
\end{align}
and
\begin{align}
\chi_{\bold n \bold m}=&4 k_{{m_\textrm{x}}}^6 \omega_{\bold n}^4 L_\textrm{x}^2 - k_{n_\perp}^4 \{-4 k_{{n_\textrm{x}}}^6 L_\textrm{x}^2 +  k_{{n_\textrm{x}}}^4 (45 -  8 \omega_{\bold m} \omega_{\bold n} L_\textrm{x}^2) \nonumber\\
&+  2 k_{{n_\textrm{x}}}^2 k_{n_\perp}^2 [51 + 6 k_{n_\perp}^2 L_\textrm{x}^2 -  8 \omega_{\bold m} \omega_{\bold n} L_\textrm{x}^2]+8 k_{n_\perp}^4 [9 + (k_{n_\perp}^2 - \omega_{\bold m} \omega_{\bold n}) L_\textrm{x}^2]\} \nonumber\\
&- 2 k_{{m_\textrm{x}}}^2 k_{n_\perp}^2 \{-4 k_{{n_\textrm{x}}}^6 L_\textrm{x}^2 + k_{n_\perp}^4 [51+ 6 k_{n_\perp}^2 L_\textrm{x}^2 - 8 \omega_{\bold m} \omega_{\bold n} L_\textrm{x}^2] \nonumber\\
&+k_{{n_\textrm{x}}}^2 k_{n_\perp}^2 [63 + 8 (k_{n_\perp}^2 -  2 \omega_{\bold m} \omega_{\bold n}) L_\textrm{x}^2] - k_{{n_\textrm{x}}}^4 [-27 + 2 (k_{n_\perp}^2 +  4 \omega_{\bold m} \omega_{\bold n}) L_\textrm{x}^2]\} \nonumber\\
&+  k_{{m_\textrm{x}}}^4 \{4 k_{{n_\textrm{x}}}^6 L_\textrm{x}^2 + k_{n_\perp}^4 (-45 + 8 \omega_{\bold m} \omega_{\bold n}L_\textrm{x}^2) + 8 k_{{n_\textrm{x}}}^4 [-3 + (k_{n_\perp}^2 + \omega_{\bold m} \omega_{\bold n}) L_\textrm{x}^2] \nonumber\\
&+2 k_{{n_\textrm{x}}}^2 k_{n_\perp}^2 [-27 +2 (k_{n_\perp}^2 +  4 \omega_{\bold m} \omega_{\bold n})L_\textrm{x}^2]\}.
\end{align}

\subsection{Quantum state evolution via master equation}\label{master}
Nonlinear interactions introduce an element of anharmonicity, thereby leading to small modifications in the energy level structure while retaining an overall quasi-harmonic character. As a result, the dynamics of the system has to be described using a generalized master equation formulated without invoking the conventional secular approximation. An appropriate theoretical framework for capturing the time evolution of the density matrix operator $\hat{\rho}$ in hybrid quantum systems subject to dissipation and thermal-like noise has been obtained in previous work   \cite{settineriDissipationThermalNoise2018}. 
The master equation reads $\dot{\hat{\rho}}=-i[\hat H,\hat{\rho}]+ \mathcal{L}\hat{\rho}$, or
\begin{align}
\label{me}
\dot{\hat{\rho}}=-i[\hat H,\hat{\rho}]+ \mathcal{L}_{\textrm{c}}\hat{\rho} + \mathcal{L}_{\textrm{w}}\hat{\rho}\;,
\end{align}
where we have split the Liouvillian superoperator $\mathcal{L}=\mathcal{L}_{\textrm{c}}+\mathcal{L}_{\textrm{w}}$ in two parts: $\mathcal{L}_{\textrm{c}}$ describes the interaction of the system with the cavity bath at temperature $T_\textrm{c}$ and with damping rate $\kappa$, whereas $\mathcal{L}_{\textrm{w}}$ describes the interaction of the system with the wall bath at temperature $T_\textrm{w}$ and with damping rate $\gamma$.

The two Liouvillian superoperators are expressed in the form:
\begin{align}
\mathcal{L}_{\textrm{c}}\hat{\rho} = \frac{\kappa}{2} \sum_{\substack{j > k \\ l > m}} &  \{  A_{lm} n_{ml}({T}) [ \hat P_{lm}  \hat{\rho} \hat P_{kj} -\hat P_{kj} \hat P_{lm} \hat{\rho}  ] + A_{kj} n_{jk}({T}) [ \hat P_{lm} \hat \rho \hat P_{kj} - \hat \rho \hat P_{kj} \hat P_{lm} ] \nonumber\\
&  +A_{lm}[n_{ml}({T}) +1] [ \hat P_{kj} \hat \rho \hat P_{lm} -\hat \rho \hat P_{lm} \hat P_{kj}]+ A_{kj}[n_{jk}({T}) +1] [ \hat P_{kj} \hat \rho \hat P_{lm} - \hat P_{lm} \hat P_{kj} \hat \rho ] \}\;.
\label{dissc}
\end{align}
and 
\begin{align}
\mathcal{L}_{\textrm{w}}\hat{\rho} = \frac{\gamma}{2} \sum_{\substack{j > k \\ l > m}}  &\{  B_{lm} n_{ml}({T}) [ \hat P_{lm}  \hat{\rho} \hat P_{kj} -\hat P_{kj} \hat P_{lm} \hat{\rho}  ]+ B_{kj} n_{jk}({T}) [ \hat P_{lm} \hat \rho \hat P_{kj} - \hat \rho \hat P_{kj} \hat P_{lm} ] \nonumber\\
&+  B_{lm}[n_{ml}({T}) +1] [ \hat P_{kj} \hat \rho \hat P_{lm} -\hat \rho \hat P_{lm} \hat P_{kj}] + B_{kj}[n_{jk}({T}) +1] [ \hat P_{kj} \hat \rho \hat P_{lm} - \hat P_{lm} \hat P_{kj} \hat \rho ] \}\;.
\label{dissw}
\end{align}
Here, the $\hat{O}_{\alpha \beta}=\langle \alpha\lvert (\hat o+\hat o^\dag)\rvert \beta\rangle$ and $\hat P_{\alpha \beta}$ define the dressed photon and phonon lowering operators $\hat{O}=\{ \hat{A}, \hat{B} \}$ in terms of their corresponding bare operators  $\hat{o}=\{ \hat{a}, \hat{b} \}$ by the relation, 
$O=\sum_{\beta,\alpha>\beta}\langle \alpha\lvert (\hat o+\hat o^\dag)\rvert \beta\rangle \hat P_{\alpha \beta}$
where the state $\rvert \alpha\rangle$ is the $\alpha$-th eigenstate of the total Hamiltonian with eigenenergy $E_\alpha$, and   $\hat P_{\alpha \beta}=\lvert \alpha\rangle\langle \beta\rvert$ indicate the projectors onto the respective eigenspaces \cite{Ridolfo2012}.
Imposing $ k_B = 1 $ for convenience, the thermal noise occupation number of the system-reservoir at real or effective temperature $ T $ is given by
\begin{equation}
n_{\alpha \beta}(T) = \left[ \exp\left( \frac{\omega_{\alpha \beta}}{T} \right) - 1 \right]^{-1}.
\label{eq:thermal_occupation}
\end{equation}
When counter-rotating terms are included in the interaction Hamiltonian it is insufficient to employ master equations in the dressed basis. In fact, a modification of the input-output relations connecting the intra-cavity field to the external fields is also necessary \cite{Macri2022,macri2022a,Mercurio2023aa}. According to these modified relations, the output fields are no longer determined by the expectation values of the bare photon operators \cite{Mercurio2022a}, but by the expectation values of the dressed operators $\hat{O}=\sum_{\beta,\alpha>\beta}\langle \alpha\lvert (\hat o+\hat o^\dag)\rvert \beta\rangle \hat P_{\alpha \beta}$.

The generalized master equation in \eqref{me} (with the dissipators defined in \eqref{dissc} and \eqref{dissw}) is not of Lindblad form, which implies that essential properties such as the positivity of the density matrix and conservation of probability are not automatically ensured. However, a detailed inspection shows that the dissipator can be regarded as approximately lindblad-like under certain conditions \cite{settineriDissipationThermalNoise2018}. Specifically, if the energy transition rate  $|E_{\beta} - E_{\alpha}|\approx |\omega_{\alpha \beta}|$ is much larger than the system damping rates $\gamma, \kappa$, the corresponding terms contribute negligibly to the dynamics. Therefore, one can assume  in the Liouvillian superoperator that $n_{jk}({T}) \approx n_{ml}({T})$ making the dissipator effectively Lindblad in nature.
Despite their limited physical relevance, these quantities result in fast-oscillating terms in the dissipative dynamics that can cause numerical instabilities and significantly increase computational cost.

To address this, we introduce a numerical filtering technique that discards terms involving large frequency mismatches
\[
\mathcal{L}^{\text{filt}} \hat{\rho} = \mathcal{L} \hat{\rho} \times F(\omega_{\alpha \beta}),
\]
with the filter function defined as $F(\omega_{\alpha \beta}) = \Theta(|\omega_{\alpha \beta}|) - \Theta(|\omega_{\alpha \beta}| - \Delta)$, where $\Theta(x)$ is the Heaviside step function and $\Delta$ is the chosen frequency cutoff bandwidth. In our numerical simulation, this filtering is used  in order to improve numerical stability and to reduce computation time while preserving the essential system dynamics.

\subsection{Quantum thermodynamics}\label{thermo}
In the last two decades, optomechanical systems have attracted a lot of interest in the context of quantum thermodynamics \cite{martinezDynamicsThermodynamicsLinear2013,vinjanampathyQuantumThermodynamics2016,strasbergFirstSecondLaw2021,pottsQuantumThermodynamics2024}, since they have been identified as possible candidates for the implementation of quantum heat engines \cite{quanQuantumThermodynamicCycles2007,quanQuantumThermodynamicCycles2009,zhangQuantumOptomechanicalHeat2014,zhangTheoryOptomechanicalQuantum2014,ianThermodynamicCycleCavity2014,mariQuantumOptomechanicalPiston2015,serafiniOptomechanicalStirlingHeat2020,PhysRevResearch.5.043274,PhysRevResearch.6.023320,hasegawa2024ultimateprecisionlimitquantum}. For this reason, a detailed quantum thermodynamic analysis of such systems is necessary to study their interaction with the environment, and therefore understand how heat flows and entropy is produced.

In the previous section, we presented a method to study the system dynamics using the generalized master equation, under the assumption that $|E_\beta - E_\alpha|\approx |\omega_{\alpha \beta}|$  much larger than the system damping rates $\gamma,\kappa$. These conditions ensure the positivity of the density matrix and the conservation of probability in numerical simulations. In order to establish an interaction with the environment we assumed that the two subsystems, namely the wall and the cavity field, are coupled to two different baths at different temperatures, $T_{\text{w}}$ and $T_{\text{c}}$ respectively, with $T_{\text{w}}>T_{\text{c}}$ \cite{kurizki2022thermodynamics}. This suggests that, if the two subsystems also interact with themselves resonantly, heat will flow from the hot bath to the cold bath, and the whole quantum system is constantly out of equilibrium \cite{kosloffQuantumThermodynamicsDynamical2013,shengNonequilibriumThermodynamicsCavity2023}. To describe the dynamics of the heat flows, we first need to recall some fundamental notion of quantum thermodynamics.

In the absence of external drives, i.e. when $\hat{H}$ is time-independent, the first law of thermodynamics in its differential form is recovered by computing the time derivative of the average energy $\langle \hat H\rangle:=\textrm{Tr}[\hat H \hat{\rho}(t)]$ as follows:
\begin{align}
\frac{d\langle \hat H\rangle}{dt}=\textrm{Tr}[\hat H\dot{\hat{\rho}}(t)]=\textrm{Tr}[\hat H\mathcal{L}\hat{\rho}(t)]\equiv \mathcal{J}(t),
\label{firstlaw}
\end{align}
where we have introduced the total heat flow $\mathcal{J}(t)=\mathcal{J}_{\text{c}}(t)+\mathcal{J}_{\text{w}}(t)$. Importantly, we define the heat-flow from the cold bath to the cavity by means of the expression, and the heat-flow from the hot bath to the wall by means of the expression. These are
\begin{subequations}\label{hf}
\begin{align}
\mathcal{J}_{\text{c}}(t)\equiv&\dot{\mathcal{Q}}_{\text{c}}(t)=\textrm{Tr}[\hat H\hat{\mathcal{L}}_{\text{c}}\hat{\rho}(t)]\\
\mathcal{J}_{\text{w}}(t)\equiv&\dot{\mathcal{Q}}_{\text{w}}(t)=\textrm{Tr}[\hat H\hat{\mathcal{L}}_{\text{w}}\hat{\rho}(t)],
\end{align}
\end{subequations}
see \cite{kosloffQuantumThermodynamicsDynamical2013, kosloffQuantumHeatEngines2014}.

Entropy production in this context is formulated in the standard manner \cite{strasbergFirstSecondLaw2021,landiIrreversibleEntropyProduction2021,PhysRevD.107.065014}, and it reads 
\begin{align}\label{epr}
\dot\Sigma=\frac{d S}{dt}-\frac{\mathcal{J}_{\text{c}}}{T_{\text{c}}}-\frac{\mathcal{J}_{\text{w}}}{T_{\text{w}}}
\end{align}
where we have introduced the von Neumann entropy $S(t)=-\textrm{Tr}[\hat{\rho}(t) \ln\hat{\rho}(t)]$ calculated with respect to density state of the system (cavity+wall) . 

Once the steady state has been reached, the flow of energy reaches a constant value and both the internal energy $\langle \hat H\rangle$ and the system entropy $S$ also become constant in time. In this case, the differential form of the first law of thermodynamics expressed in \eqref{firstlaw} reduces to $\mathcal{J}_{\text{c}}+\mathcal{J}_{\text{w}}=0$, suggesting that the dynamics is essentially governed by the heat flows between the two baths. In this scenario, the entropy production rate becomes $\dot\Sigma=-\mathcal{J}_{\text{c}}/T_{\text{c}}-\mathcal{J}_{\text{w}}/T_{\text{w}}=\mathcal{J}_{\text{w}}(1/T_{\text{c}}-1/T_{\text{w}})$, which is clearly always positive, since $\mathcal{J}_{\text{w}}>0$ and $T_{\text{w}}>T_{\text{c}}$.

The non-negativity of entropy production can be demonstrated even beyond the steady-state regime. 
To see this we note that the time derivative of the von Neumann entropy is given by
\begin{align}
    \frac{dS}{dt} &= -\mathrm{Tr}\left(\frac{d\hat\rho}{dt}\ln\hat\rho\right)
    = -\mathrm{Tr}\left[(\mathcal{L}\hat\rho)\ln\hat\rho\right]= -\mathrm{Tr}[\mathcal{L}_{\mathrm{w}}\hat\rho\ln\hat\rho]-\mathrm{Tr}[\mathcal{L}_{\mathrm{c}}\hat\rho\ln\hat\rho],
    \label{eq:dSdt}
\end{align}
where $\mathcal{L}=\mathcal{L}_{\mathrm{w}}+\mathcal{L}_{\mathrm{c}}$.
The entropy production rate is then expressed as
\begin{align}
    \dot{\Sigma} &= -\mathrm{Tr}[\mathcal{L}_{\mathrm{w}}\hat\rho\ln\hat\rho]
    -\mathrm{Tr}[\mathcal{L}_{\mathrm{c}}\hat\rho\ln\hat\rho]-\frac{\mathrm{Tr}(H\mathcal{L}_{\mathrm{w}}\hat\rho)}{T_{\mathrm{w}}}
    -\frac{\mathrm{Tr}(H\mathcal{L}_{\mathrm{c}}\hat\rho)}{T_{\mathrm{c}}}.
    \label{eq:Sigma_dot}
\end{align}
The Spohn inequality \cite{Spohn1978entropyproduction} states that, for arbitrary positive density operators $\hat\rho$ and $\hat\sigma$, the following relation holds:
\begin{align}
    -\mathrm{Tr}[\mathcal{L}_\alpha\hat\rho(\ln\hat\rho-\ln\hat\sigma)]&\ge0,
    \label{eq:Spohn_inequality1}
\end{align}
where $\alpha = \mathrm{c}$ or $\mathrm{w}$. 


To prove the non-negativity of entropy production, in \eqref{eq:Spohn_inequality1} we assume that the each bath is found in a Gibbs state of corresponding temperature. Concretely, we impose $\hat\sigma=\hat\rho_\alpha^{\mathrm{eq}}$, where $ \hat\rho_\alpha^{\mathrm{eq}} = e^{-H / T_\alpha}/Z_\alpha$ and $\quad Z_\alpha = \operatorname{Tr}(e^{-H / T_\alpha})$.
Then, the entropy production rate $\dot{\Sigma}$ is guaranteed to be non-negative at all times ($\dot{\Sigma} \geq 0$), thereby confirming the second law of thermodynamics.

\section{Modified perturbation theory}\label{perthe}
The interaction Hamiltonian in \eqref{quantum:hamiltonian:explicit} consists of the sum of three parts proportional to increasing power of $\epsilon$. If we assume that $\epsilon$ is the intensity of the perturbation of our Hamiltonian, the application of the perturbation theory must account for the fact that the power of the perturbation is not uniform in the interaction Hamiltonian.
In this section, we modify the perturbed eigenstates of the Hamiltonian up to the second order, and the eigenenergies up to the third order, taking into account the fact that $H_1$, $H_2$ and $H_3$ are proportional to $\epsilon$, $\epsilon^2$ and $\epsilon^3$, respectively. For the sake of simplicity, we consider a non-degenerate, time-independent Hamiltonian. This means that, for the purposes of the current analysis, we do not need to activate any of the possible resonant conditions between the mechanical and the optical degrees of freedom.

We start from a Hamiltonian that can be split as usual into two parts, $\hat H=\hat H_0+\hat H_\textrm{I}$, where the sets of eigenstates and eigenenergies of the unperturbed Hamiltonian $\hat H_0$ are henceforth designed by $\lvert n^{(0)}\rangle$ and $E_n^{(0)}$, respectively. We assume that the perturbed Hamiltonian consists of three terms of the form $\hat H_\textrm{I}=\epsilon \hat H_1+\epsilon^2 \hat H_2+\epsilon^3 \hat H_3$. The complete set of eigenstates of the total Hamiltonian, as well as their relative eigenenergies, can be expressed in power series
\begin{align}
\lvert n\rangle=&\lvert n^{(0)}\rangle+\epsilon \lvert n^{(1)}\rangle+\epsilon^2 \lvert n^{(2)}\rangle+\epsilon^3 \lvert n^{(3)}\rangle+...\\
E_n=& E_n^{(0)}+\epsilon E_n^{(1)}+\epsilon^2 E_n^{(2)}+\epsilon^3 E_n^{(3)}+...
\end{align}
where $\lvert n^{(j)}\rangle$ and $E_n^{(j)}$ are the $j$-order corrections to the eigenstates and the eigenenergy, respectively.

The time-independent Schrödinger equation therefore reads
\begin{align}
\left(\hat H_0+\epsilon \hat H_1+\epsilon^2 \hat H_2+\epsilon^3 \hat H_3\right)\left(\lvert n^{(0)}\rangle+\epsilon \lvert n^{(1)}\rangle+\epsilon^2 \lvert n^{(2)}\rangle+\epsilon^3 \lvert n^{(3)}\rangle+...\right)\nonumber\\
=\left(E_n^{(0)}+\epsilon E_n^{(1)}+\epsilon^2 E_n^{(2)}+\epsilon^3 E_n^{(3)}+...\right)\left(\lvert n^{(0)}\rangle+\epsilon \lvert n^{(1)}\rangle+\epsilon^2 \lvert n^{(2)}\rangle+\epsilon^3 \lvert n^{(3)}\rangle+...\right).
\label{SE}
\end{align}
We now provide a solution for the Schrödinger equation at each order in $\epsilon$, exploiting the fact that at the 0-order it is already solved: $\hat H_0\lvert n^{(0)}\rangle=E_n^{(0)}\lvert n^{(0)}\rangle$.
\subsection{First order in $\epsilon$}
The first order corrections to the Schrödinger equation does not differ from what we can find in any textbook \cite{sakurai2020modern}, but it is here reported for completeness, and because it provides the general technique to calculate the higher orders. We first collect all terms in Eq.\ref{SE} proportional in $\epsilon$:
\begin{align}
\hat H_0\lvert n^{(1)}\rangle+\hat H_1\lvert n^{(0)}\rangle=E_n^{(0)}\lvert n^{(1)}\rangle+E_n^{(1)}\lvert n^{(0)}\rangle.
\label{order1}
\end{align}
By applying the standard procedure in perturbation theory, we multiply by the bra $\langle n^{(0)}\rvert$ on the left, and use the fact that $E_n^{(0)}$ is the eigenvalue of the Hamiltonian $\hat H_0$. We therefore obtain the well-known result
\begin{align}
E_n^{(1)}=\langle n^{(0)}\lvert\hat H_1\rvert n^{(0)}\rangle 
\label{E1}
\end{align}
The state $\lvert n^{(1)}\rangle$ can always be decomposed with respect to the complete set $\lvert n^{(0)}\rangle$ as $\lvert n^{(1)}\rangle=\sum_{m\neq n} c_{nm}^{(1)} \lvert m^{(0)}\rangle$, where $c_{nm}^{(1)}$ are the coefficients of the expansion. To achieve the explicit expression of these coefficients, we insert the decomposition of $\lvert n^{(1)}\rangle$ written above into \eqref{order1}, and multiply by $\langle l^{(0)}\rvert$ on the left. We obtain 
\begin{align}
\sum_{m\neq n} c_{nm}^{(1)}\langle l^{(0)}\rvert \hat H_0\lvert m^{(0)}\rangle+\langle l^{(0)}\rvert\hat H_1\lvert n^{(0)}\rangle=\sum_{m\neq n} c_{nm}^{(1)}E_n^{(0)}\langle l^{(0)}\mid m^{(0)}\rangle+E_n^{(1)}\langle l^{(0)}\mid n^{(0)}\rangle.
\end{align}
This equation recovers \eqref{E1} if $l=n$. If $l\neq n$ we have
\begin{align}
c_{nl}^{(1)}E_l^{(0)}+\langle l^{(0)}\rvert\hat H_1\lvert n^{(0)}\rangle=c_{nl}^{(1)}E_n^{(0)},
\end{align}
from which we can isolate the coefficients
\begin{align}
c_{nl}^{(1)}=\frac{\langle l^{(0)}\rvert\hat H_1\lvert n^{(0)}\rangle}{E_n^{(0)}-E_l^{(0)}}
\end{align}
and finally obtain the first order correction to the Hamiltonian eigenstates:
\begin{align}
\lvert n^{(1)}\rangle=\sum_{l\neq n} \frac{\langle l^{(0)}\rvert\hat H_1\lvert n^{(0)}\rangle}{E_n^{(0)}-E_l^{(0)}}\lvert l^{(0)}\rangle.
\label{n1}
\end{align}
\subsection{Second order in $\epsilon$}\label{sec:p1}
We now use the same technique utilized above to calculate the corrections at the second order in $\epsilon$ to both the eigenstates and the eigenenergies of the Hamiltonian. We therefore start from the second-order Schrödinger equation
\begin{align}
\hat H_0 \lvert n^{(2)}\rangle+\hat H_1 \lvert n^{(1)}\rangle+\hat H_2\lvert n^{(0)}\rangle=E_n^{(0)}\lvert n^{(2)}\rangle+E_n^{(1)}\lvert n^{(1)}\rangle+E_n^{(2)}\lvert n^{(0)}\rangle.
\label{order2}
\end{align}
Note that the last term on the left side is normally not present in the standard perturbation theory. The correction to the energy levels are again given by multiplying by $\langle n^{(0)}\rvert$ on the left, and exploiting the zeroth-order Schrödinger equation. This time, we get
\begin{align}
\langle n^{(0)}\rvert \hat H_1 \lvert n^{(1)}\rangle+\langle n^{(0)}\rvert \hat H_2 \lvert n^{(0)}\rangle=E_n^{(1)}\langle n^{(0)}\mid n^{(1)}\rangle+E_n^{(2)}=E_n^{(2)},
\end{align}
where the last equivalence stems from the orthogonality between $\lvert n^{(1)}\rangle$ and $\lvert n^{(0)}\rangle$. The second-order correction to the Hamiltonian eigenenergies is therefore given by
\begin{align}
E_n^{(2)}=\langle n^{(0)}\rvert \hat H_1 \lvert n^{(1)}\rangle+\langle n^{(0)}\rvert \hat H_2 \lvert n^{(0)}\rangle=\sum_{l\neq n} \frac{\lvert\langle l^{(0)}\rvert\hat H_1\lvert n^{(0)}\rangle\rvert^2}{E_n^{(0)}-E_l^{(0)}}+\langle n^{(0)}\rvert \hat H_2 \lvert n^{(0)}\rangle,
\end{align}
where in the last equivalence we substitute the explicit expression of $\lvert n^{(1)}\rangle$ already calculated in \eqref{n1}. We notice that the presence of the last term differs from the ordinary perturbation theory. Interestingly, as the first order correction depends on the average value of $\hat H_1$ with respect to the unperturbed basis, the presence of a perturbation proportional to $\epsilon^2$ induces a correction at the second order which is again equal to the average value of $\hat H_2$ with respect to the unperturbed basis.

To calculate the correction to the eigenstates, we employ the same procedure as Sec.\ref{sec:p1}. Each eigenstates $\lvert n^{(2)}\rangle$ can be expressed as a linear combination of unperturbed states:
\begin{align}
\lvert n^{(2)}\rangle=\sum_{m\neq n} c_{nm}^{(2)} \lvert m^{(0)}\rangle,    
\label{dec2}
\end{align}
where $c_{nm}^{(2)}$ is the set of coefficients of the linear decomposition. To calculate the form of these coefficient, we replace \eqref{dec2} into \eqref{order2} and multiply by $\langle l^{(0)}\rvert$ on the left, thereby obtaining
\begin{align}
 \sum_{m\neq n} c_{nm}^{(2)} \hat H_0\lvert m^{(0)}\rangle+\hat H_1 \lvert n^{(1)}\rangle+\hat H_2\lvert n^{(0)}\rangle=E_n^{(0)}\sum_{m\neq n} c_{nm}^{(2)} \lvert m^{(0)}\rangle+E_n^{(1)}\lvert n^{(1)}\rangle+E_n^{(2)}\lvert n^{(0)}\rangle.
\end{align}
Note that we already know the explicit expression for $\lvert n^{(1)}\rangle$. Substituting \eqref{n1} into the equation written above, we get
\begin{align}
 \sum_{m\neq n} c_{nm}^{(2)} (E_n^{(0)}-\hat H_0)\lvert m^{(0)}\rangle+ \sum_{m\neq n} \frac{\langle m^{(0)}\rvert\hat H_1\lvert n^{(0)}\rangle}{E_n^{(0)}-E_m^{(0)}}(E_n^{(1)}-\hat H_1)\lvert m^{(0)}\rangle+(E_n^{(2)}-\hat H_2)\lvert n^{(0)}\rangle=0.
\end{align}
As usual, we multiply by the bra $\langle l^{(0)}\rvert$ on the left,
\begin{align}
c_{nl}^{(2)} (E_n^{(0)}-E_l^{(0)})+ \sum_{m\neq n} \frac{\langle m^{(0)}\rvert\hat H_1\lvert n^{(0)}\rangle}{E_n^{(0)}-E_m^{(0)}}\left(E_n^{(1)}-\langle l^{(0)}\rvert\hat H_1\lvert m^{(0)}\rangle\right)-\langle l^{(0)}\rvert\hat H_2\lvert n^{(0)}\rangle=0,
\end{align}
and isolate the coefficient from the rest
\begin{align}
c_{nl}^{(2)} = \sum_{m\neq n} \frac{\langle l^{(0)}\rvert\hat H_1\lvert m^{(0)}\rangle}{E_n^{(0)}-E_l^{(0)}}\frac{\langle m^{(0)}\rvert\hat H_1\lvert n^{(0)}\rangle}{E_n^{(0)}-E_m^{(0)}}-\frac{\langle l^{(0)}\rvert\hat H_1\lvert n^{(0)}\rangle\langle n^{(0)}\lvert\hat H_1\rvert n^{(0)}\rangle}{\left(E_n^{(0)}-E_l^{(0)}\right)^2}+\frac{\langle l^{(0)}\rvert\hat H_2\lvert n^{(0)}\rangle}{E_n^{(0)}-E_l^{(0)}}.
\end{align}
Note that in the formula above we wrote $E_n^{(1)}$ explicitly via \eqref{E1}. The second-order correction to the state is finally given by
\begin{align}
\lvert n^{(2)}\rangle=\sum_{l\neq n}\left[\sum_{m\neq n} \frac{\langle l^{(0)}\rvert\hat H_1\lvert m^{(0)}\rangle}{E_n^{(0)}-E_l^{(0)}}\frac{\langle m^{(0)}\rvert\hat H_1\lvert n^{(0)}\rangle}{E_n^{(0)}-E_m^{(0)}}-\frac{\langle l^{(0)}\rvert\hat H_1\lvert n^{(0)}\rangle\langle n^{(0)}\lvert\hat H_1\rvert n^{(0)}\rangle}{\left(E_n^{(0)}-E_l^{(0)}\right)^2}+\frac{\langle l^{(0)}\rvert\hat H_2\lvert n^{(0)}\rangle}{E_n^{(0)}-E_l^{(0)}}\right] \lvert l^{(0)}\rangle,    
\label{n2}
\end{align}
This formula differs from the standard one found in literature \cite{PhysRevLett.111.060403} because of the last contribution, which determines the transition between two eigenstates of the unperturbed Hamiltonian due to the second-order interaction Hamiltonian $\hat H_2$.

Before proceeding with the third order corrections, which will be carried out only to the eigenenergies, we remind that that corrected state up to the second order still must be renormalized. We renormalize the eigenstate as $\lvert \tilde n \rangle=C^{1/2}\lvert  n \rangle$, where the factor $C$ ensures that $\langle \tilde n\mid \tilde n\rangle=1$. We notice that the calculation of the normalization factor at the second order in $\epsilon$ depends only on the correction at the first order $\lvert n^{(1)}\rangle$, and it is therefore easy to find in literature \cite{}. However, it is reported here for completeness:
\begin{align}
C=1-\epsilon^2\sum_{l\neq n}\frac{\lvert\langle l^{(0)}\rvert\hat H_1\lvert n^{(0)}\rangle\rvert^2}{\left(E_n^{(0)}-E_l^{(0)}\right)^2}.
\end{align}

\subsection{Third order in $\epsilon$}
We conclude this section by reporting the correction to the eigenenergies at the third order in $\epsilon$. If we isolate all terms proportional to $\epsilon^3$ in \eqref{SE} we obtain
\begin{align}
\hat H_0\lvert n^{(3)}\rangle+\hat H_2\lvert n^{(1)}\rangle+\hat H_1\lvert n^{(2)}\rangle+\hat H_3\lvert n^{(0)}\rangle=E_n^{(0)}\lvert n^{(3)}\rangle+E_n^{(2)}\lvert n^{(1)}\rangle+E_n^{(1)}\lvert n^{(2)}\rangle+ E_n^{(3)}\lvert n^{(0)}\rangle.
\label{order3}
\end{align}
By multiplying each term by the usual $\langle n^{(0)}\rvert$ and exploit the orthogonality between $\lvert n^{(0)}\rangle$ and its high order corrections, we finally get
\begin{align}
E_n^{(3)}=&\langle n^{(0)}\rvert \hat H_2\lvert n^{(1)}\rangle+\langle n^{(0)}\rvert \hat H_1\lvert n^{(2)}\rangle+\langle n^{(0)}\rvert \hat H_3\lvert n^{(0)}\rangle\nonumber\\
=&\sum_{l\neq n}\left[\sum_{m\neq n} \frac{\langle n^{(0)}\rvert\hat H_1\lvert m^{(0)}\rangle}{E_n^{(0)}-E_m^{(0)}}\frac{\langle m^{(0)}\rvert\hat H_1\lvert l^{(0)}\rangle}{E_n^{(0)}-E_l^{(0)}}-\frac{\langle n^{(0)}\rvert\hat H_1\lvert n^{(0)}\rangle\langle n^{(0)}\lvert\hat H_1\rvert l^{(0)}\rangle}{\left(E_n^{(0)}-E_l^{(0)}\right)^2}+2\frac{\langle n^{(0)}\rvert\hat H_2\lvert l^{(0)}\rangle}{E_n^{(0)}-E_l^{(0)}}\right] \langle l^{(0)}\rvert \hat H_1\lvert n^{(0)}\rangle\nonumber\\
&+\langle n^{(0)}\rvert \hat H_3\lvert n^{(0)}\rangle.
\end{align}
\end{document}